\documentclass[aps,showpacs,eqsecnum,10pt]{revtex4}

\usepackage{graphicx}
\usepackage{amsmath}
\usepackage{amssymb}

\newcommand{\be}{\begin{equation}}
\newcommand{\ee}{\end{equation}}
\newcommand{\ba}{\begin{eqnarray}}
\newcommand{\ea}{\end{eqnarray}}

\newcount\bozza \bozza=0
\ifnum\bozza=1
\newdimen\shift \shift=-2truecm
\def\lb#1{%
{\label{#1}\rlap{\kern\shift{$\scriptstyle#1$}}}}
\else\def\lb#1{\label{#1}} \fi

\begin{document}


\title{Transport of Dirac quasiparticles in graphene: Hall and optical conductivities}

\author{V.P.~Gusynin$^{1}$}
\email{vgusynin@bitp.kiev.ua}
\author{S.G.~Sharapov$^{1,2}$}
\email{sharapov@bitp.kiev.ua}

\affiliation{$^1$ Bogolyubov Institute for Theoretical Physics,
        Metrologicheskaya Str. 14-b, Kiev, 03143, Ukraine\\
        $^2$ Department of Physics and Astronomy, McMaster University,
        Hamilton, Ontario, Canada, L8S 4M1}

\date{\today }

\begin{abstract}
The analytical expressions for both diagonal and off-diagonal ac
and dc conductivities of graphene placed in an external magnetic
field are derived. These conductivities exhibit rather unusual
behavior as functions of frequency, chemical potential and applied
field which is caused by the fact that the quasiparticle
excitations in graphene are Dirac-like. One of the most striking
effects observed in graphene is the odd integer quantum Hall
effect. We argue that it is caused by the anomalous properties of
the Dirac quasiparticles from the lowest Landau level.  Other
quantities such as Hall angle and Nernst signal also exhibit
rather unusual behavior, in particular when there is an excitonic
gap  in the spectrum of the Dirac quasiparticle excitations.
\end{abstract}




\maketitle

\section{Introduction}

There is significant progress in fabrication of free-standing
monocrystalline graphite films with thickness down to a single
atomic layer \cite{Novoselov2004Science,Novoselov2005PNAS} and the
relatively thick (thicker than 3 monolayers) graphite films
\cite{Berger2004JPCB,Zhang2005PRL,Bunch2005Nano,Morozov2005} are now
widely produced. The new one layer material, called graphene,
possesses truly remarkable properties both from a technological and
theoretical point of view. Graphene is a promising candidate for
applications in future micro- and nanoelectronics due to its
excellent mechanical characteristics, scalability to the nanometer
sizes, and the ability to sustain huge ($>10^8 A/cm^2$) electric
currents. By using the electric field effect
\cite{Novoselov2004Science,Novoselov2005PNAS,Zhang2005PRL,Bunch2005Nano,Morozov2005,Geim-new,Kim-new},
it is possible to change the carrier concentration in samples by
tens times and even to change the carrier type from electron to hole
when the sign of applied gate voltage is reversed. Another interest
in graphene is related to the fact that it represents a building
block for the other forms of carbon, viz. graphite is a stack of
graphene layers, carbon nanotubes are wrapped graphene layers, while
fullerenes can be created from graphene by introducing topological
defects.

On the theoretical side, the conduction and valence bands in
graphene touch upon each other at isolated points in the Brillouin
zone and this results in the linear, Dirac-like (up to energies of
the order of $1000$\,K) spectrum of quasiparticle excitations which
makes graphene a unique truly two-dimensional "relativistic"
electronic system.  The thinnest graphite films can be described by
a low-energy (2+1) dimensional effective {\em massless\/} Dirac
theory \cite{Semenoff1984PRL,Gonzales1993NP}. A recent observation
\cite{Geim-new,Kim-new} of the unconventional integer quantum Hall
effect (IQHE)
\begin{equation}
\label{Hall-Dirac} \sigma_{xy} = - \frac{2e^2}{h}(2n+1), \qquad n
=0,1,\ldots
\end{equation}
which is expected  from the analytical study  \cite{Gusynin2005PRL}
based on the fundamental properties of the $2+1$ dimensional Dirac
theory, can be considered as the ultimate proof of the existence of
the Dirac quasiparticles in this fascinating material. A
complementary numerical investigation of the Landau level structure
for a hexagonal lattice model with the nearest neighbor and
next-nearest neighbor hoppings also led the authors of
Ref.~\cite{Peres2005} to the conclusion that the Hall conductivity
is quantized according to the rule (\ref{Hall-Dirac}). In contrast
to this behavior expected for an ideal 2D graphene, thicker 2 to 10
layers thick films studied in
Refs.~\cite{Novoselov2004Science,Novoselov2005PNAS,Berger2004JPCB}
exhibit instead a conventional Hall quantization $\sigma_{xy} = - 4
(e^2/h) n$.

The Dirac quasiparticles seem to be present not only in graphene,
but also in the highly oriented pyrolitic graphite (HOPG), single
crystalline Kish graphite and the relatively thick (thicker than
3-10 monolayers) graphite films, where warping introduces other
types of carriers \cite{Dresselhaus1974PRB}. The Hall effect
features in HOPG graphite were observed in
Refs.~\cite{Uji1998PB,Kopelevich2003PRL} (see also the latest
Ref.~\cite{Kempa2006SSC}), but the Hall conductivity quantization in
these systems remains conventional
\cite{Novoselov2004Science,Berger2004JPCB,Ocana2003PRB,Kopelevich2003PRL}.
Nevertheless, the presence of the Dirac quasiparticles can be
detected using other experimental techniques. For example,  the
differences between the Dirac and Schr\"odinger (massive)
quasiparticles may be observed in thermodynamic and magnetotransport
measurements
\cite{Novoselov2004Science,Ocana2003PRB,Kopelevich2003PRL,Berger2004JPCB,Zhang2005PRL,Bunch2005Nano,Morozov2005}.
For instance, the phase of de Haas van Alphen and Shubnikov de Haas
oscillations for Dirac quasiparticles is shifted by $\pi$
\cite{Sharapov2004PRB,Gusynin2005PRB,Luk'yanchuk2004PRL,Geim-new,Kim-new}
compared to the phase of non-relativistic quasiparticles. Moreover,
the Dingle and temperature factors in the amplitude of oscillations
explicitly depend on the carrier density in the case of a Dirac-like
spectrum \cite{Sharapov2004PRB,Gusynin2005PRB}. These two
characteristic features allow one to distinguish the Dirac
quasiparticles from very light, $\sim 0.01 m_e$ ($m_e$ is the
electron mass) particles, and from other carriers which are also
present in graphite \cite{Luk'yanchuk2004PRL}. The Landau levels in
graphite are  observed in high magnetic field using both scanning
tunneling spectroscopy \cite{Matsui2005PRL} and infrared
spectroscopy \cite{Li2005}. The latter allowed one to observe in
HOPG the cyclotron resonance modes and to establish that some of
them reveal a $\sqrt{B}$ dependence of the cyclotron frequency which
is expected for the Dirac quasiparticles. Actually, this
characteristic, $\sqrt{B}$, dependence shows up in various
properties related to the Dirac quasiparticles. An interesting
example is the magnetization $M \sim - \sqrt{B}$ (cf. Eq.~(7.4) of
Ref.~\cite{Sharapov2004PRB}) at zero chemical potential $\mu$ which
implies that the magnetic susceptibility $\chi =
\partial M /\partial B \varpropto -B^{-1/2}$ diverges at zero field.
Although the singularity of $\chi(B \to 0)$ is smoothed [see
Ref.~\cite{Nersesyan1989JLTP}, where another system with Dirac
quasiparticles is considered] by a coupling between layers, finite
temperature and/or chemical potential, the presence of Dirac
quasiparticles results in an anomalously strong diamagnetism of
graphite \cite{Beaugnon1991Nature}.

The purpose of the present work is to extend the analysis made in
the previous papers
\cite{Gorbar2002PRB,Sharapov2004PRB,Gusynin2005PRB} (see also
Ref.~\cite{Sharapov2003PRB} devoted to a so called $d$-density wave
state which is also described by the same low-energy Dirac
Lagrangian) where thermodynamic and mostly diagonal dc
magnetotransport properties of graphene were studied. We derive
analytical expressions both for diagonal and off-diagonal ac
conductivity which in contrast to the previous papers include a {\em
frequency dependent} impurity scattering rate. Then we concentrate
mostly on the dc Hall conductivity giving throughout derivations of
the results presented in our short paper Ref.~\cite{Gusynin2005PRL}
and considering the limiting cases that were not yet considered.

The paper is organized as follows. In Sec.~\ref{sec:model} general
features of the model for a single layer of graphene are
described. In Sec.~\ref{sec:conductivity-general} we present the
analytical expressions for both diagonal and off-diagonal ac
conductivities, including dc limits of these expressions (the
details of calculation are given in Appendix~\ref{sec:A}). Then in
Sec.~\ref{sec:Hall} we consider the dc Hall conductivity, the Hall
angle and the Nernst signal are studied in Sec.~\ref{sec:Nernst}.
In particular in Sec.~\ref{sec:detection} we discuss a possibility
of detecting a gap that may exist in the spectrum of the
quasiparticle excitations of graphene. In Conclusions,
Sec.~\ref{sec:concl} we give a concise summary of the obtained
results. The extra technical details concerning the Hall
conductivity in the clean limit are given in Appendix~\ref{sec:B}.
The equation for chemical potential $\mu$ is considered in
Appendix~\ref{sec:D} and the solution of the Dirac equation in the
symmetric gauge is presented in Appendix~\ref{sec:E}.

\section{Model}
\label{sec:model}

As discussed, for example, in
Refs.~\cite{Gorbar2002PRB,Sharapov2004PRB,Gusynin2005PRB},
we start directly from the conventional QED$_{2+1}$ Lagrangian density
\begin{equation}
\label{Lagrangian} \mathcal{L} = \sum_{\sigma= \pm 1}
\bar{\Psi}_{\sigma} (t, \mathbf{r}) \left[ i \gamma^0 (\hbar
\partial_t - i \mu_\sigma) + i v_F \gamma^1 \left(\hbar \partial_x + i
\frac{e}{c}A_x^{\mathrm{ext}}\right) + i v_F \gamma^2 \left(\hbar
\partial_y + i \frac{e}{c}A_y^{\mathrm{ext}}\right) - \Delta \right]
\Psi_{\sigma}(t, \mathbf{r}),
\end{equation}
where $\Psi_{\sigma} = (\psi_{1\sigma}(t, \mathbf{r}), \psi_{2
\sigma}(t, \mathbf{r}))$ is the four-component Dirac spinor combined
from two spinors $\psi_{1\sigma}, \psi_{2 \sigma}$ [corresponding to
$\mathbf{K}$ and $\mathbf{K}^\prime$ points of the Fermi surface,
respectively] that describe the Bloch states residing on the two
different sublattices of the biparticle hexagonal lattice of the
graphene sheet.  In Eq.~(\ref{Lagrangian}) $\gamma^\nu$ with
$\nu=0,1,2$ are $4\times
4$ $\gamma$ matrices belonging to a reducible representation in
$2+1$, for example, $\gamma^\nu = \sigma_3\otimes (\sigma_3, i
\sigma_2, - i \sigma_1)$, $\bar{\Psi}_\sigma = \Psi_{\sigma}^\dagger
\gamma^0$ is the Dirac conjugated spinor, $-e<0$ is the  electron
charge, $v_F$ is the Fermi velocity, and $\sigma = \pm 1$ is the
spin variable. More generally the number of spin components $N_f$
can be regarded as a flavor index and $N_f=2$ corresponds to the
physical case.

The external magnetic field $\mathbf{B}= \nabla \times
\mathbf{A}^{\mathrm{ext}}$ is applied perpendicular to the plane
along the positive z axis and the corresponding vector potential
is taken in the symmetric gauge $\mathbf{A}^{\mathrm{ext}} =
(-By/2, Bx/2)$. The energy scale associated with the magnetic
field expressed in the units of temperature reads
\begin{equation}
\frac{e B v_F^2}{c} \to \frac{eB \hbar v_F^2}{c}
\frac{1}{k_B^2}(K^2) = 8.85 \times 10^{-8} v_F^2(\mbox{m/s}) B(T),
\end{equation}
where $v_F$ and $B$ are given in m/s and Tesla, respectively.  In
the following we set $\hbar = k_B=1$, and in some places $e=c=1$,
unless stated explicitly otherwise. There is some disagreement in
the literature concerning the precise value of $v_F$ in graphene
which is related to an uncertainty in the value of the
nearest-neighbor hopping $t$. For  numerical calculations we assume
that $t \sim 2.3 \mbox{eV}$, so that $v_F \approx 7.4 \times 10^5
\mbox{m/s}$ which leads to the relationship $eB \to (4.85 \times
10^4 \mbox{K}^2) B(\mbox{T})$. Note that the latest experiments
\cite{Geim-new,Kim-new} indicate that $v_{F} \approx (1 \div 1.1)
\times 10^6 \mbox{m/s}$.

Since the Lagrangian (\ref{Lagrangian}) originates from a
nonrelativistic many-body theory, the Zeeman interaction term has to
be explicitly included by considering spin splitting $\mu_{\sigma} =
\mu - \sigma g/2 \mu_B B$ of the chemical potential $\mu$, where
$\mu_B = e \hbar/(2mc)$ is the Bohr magneton and $g$ is the Lande
factor. However, for the relativistic quasiparticle spectrum with
the realistic values of $v_F \varpropto 10^6 \mbox{m/s}$ and $g \sim
2$ the distance between Landau levels turns out to be very large
compared to the Zeeman splitting \cite{Gusynin2005PRB}, so that in
what follows we will not consider this term and just multiply all
relevant expressions by the above-mentioned number of flavors $N_f$.
We note, however, that the latest measurements in high fields (up to
$45 \mbox{T}$) \cite{Zhang2006} revealed a lifting of the spin and
sublattice degeneracy, so that the half integer Hall quantization
changes to the integer one for fields $B>20 \mbox{T}$.

While simple tight-binding calculations (see e.g.
Ref.~\cite{Saito.book}) made for hexagonal lattice of a single
graphene sheet predict that $\mu =0$, the real picture is more
complicated and the actual value of $\mu$ in HOPG is nonzero due to
inter-layer hopping, finite doping, and/or disorder. Moreover, a
nonzero and even tunable value of $\mu$ (including the change of the
character of carriers, either electron or holes) is possible in the
electric-field doping experiments made on monocrystalline graphitic
films
\cite{Novoselov2004Science,Novoselov2005PNAS,Zhang2005PRL,Bunch2005Nano,Geim-new,Kim-new,Morozov2005}.
In our notations $\mu>0$ corresponds to electrons and, accordingly,
to the positive gate voltage, $V_g$.

The Lagrangian (\ref{Lagrangian}) also includes a gap $\Delta$, so
that for $B=0$ it describes quasiparticles with the dispersion
$E(\mathbf{k}) = -\mu \pm \sqrt{v_F^2 \mathbf{k}^2 + \Delta^2}$.
Again, this gap is zero when non-interacting quasiparticles on the
hexagonal lattice with nearest neighbor hopping are considered.
However, it could open as a result of poor screening of the Coulomb
interaction in graphite \cite{Khveshchenko2001PRL,Gorbar2002PRB}
and/or in the presence of an external magnetic field (the phenomenon
of magnetic catalysis) \cite{Gusynin1995PRD}. The physical meaning
of this gap (or a singlet excitonic order parameter) is directly
related to the electron density imbalance between the A and B
sublattices of the bi-particle hexagonal lattice of graphene
\cite{Khveshchenko2001PRL}. The opening of such a gap was already
the subject of an experimental investigation
\cite{Kopelevich2003PRL} and we hope that the predictions made in
Refs.~\cite{Khveshchenko2001PRL,Gorbar2002PRB} will be tested again
on the new thin samples that are closer to the ideal graphene
considered in these theoretical papers.

In contrast to the diagonal transport coefficients, the off-diagonal
transport properties are sensitive to the sign of the product $e B$.
Thus for the lucidity of the presentation, we begin with the
expression for the spectral function of Dirac fermions and perform
the calculation without assuming the positiveness of the product
$eB$.

The Green's function of Dirac fermions described by the Lagrangian
(\ref{Lagrangian}) in an external field reads
\begin{equation}
\label{Green} S(t-t^\prime, \mathbf{r};\mathbf{r}^\prime)=\exp
\left( -\frac{ie}{c}
\mathbf{r}\mathbf{A}^{ext}(\mathbf{r^\prime})\right)
\tilde{S}(t-t^\prime, \mathbf{r}-\mathbf{r}^\prime),
\end{equation}
where $\tilde{S}(t-t^\prime, \mathbf{r}-\mathbf{r}^\prime)$ is the
translation invariant part of $S(t-t^\prime,
\mathbf{r}-\mathbf{r}^\prime)$. Its derivation using the Schwinger
proper-time method and decomposition over Landau level poles has
been discussed in many papers (see, e.g. Refs.
\cite{Gusynin1995PRD,Chodos1990PRD,Sharapov2003PRB}), so here we
begin with the Fourier transform of ${\tilde S}(x-y)$ in the
Matsubara representation
\be
S(i\omega_m,\mathbf{k})=e^{-\frac{c{\mathbf{k}}^{2}}{|eB|}}\sum\limits_{n=0}^\infty
(-1)^n\frac{S_n(i\omega_m,\mathbf{k})}{(i\omega_m)^2-M_n^2},
\qquad \omega_m = \pi (2m+1) T, \ee
where $T$ is the temperature,
\begin{equation}
\label{Mn} M_{n}=\sqrt{\Delta^{2}+ 2 nv_F^2|eB|/c}
\end{equation}
are the energies of the relativistic Landau levels and
\be \label{S_n}
S_n(i\omega_m,\mathbf{k})=2(i\omega_m\gamma^0+\Delta)\left[P_{-}L_n\left(\frac{2c\mathbf{k}^2}{|eB|}
\right)-P_{+}L_{n-1}\left(\frac{2c\mathbf{k}^2}{|eB|}\right)\right]-4\mathbf{k}{\pmb{\gamma}}
L_{n-1}^1\left(\frac{2c\mathbf{k}^2}{|eB|}\right),\ee
with $P_{\pm }=(1\pm i\gamma ^{1}\gamma ^{2}{\rm sgn}(eB))/2$ being
projectors and $L^{\alpha}_{n}(z)$ the generalized Laguerre
polynomials. By definition, $L_{n}(z)\equiv L^{0}_{n}(z)$ and
$L^{\alpha}_{-1}(z)\equiv 0$.

In what follows we also need the retarded and advanced Green's
functions that are obtained by analytic continuation from positive
and negative discrete frequencies, respectively,
$S^R(\omega+i0,\mathbf{k})=S(i\omega_m \to \omega+i0,\mathbf{k})$
and $S^A(\omega-i0,\mathbf{k})=S(i\omega_m \to
\omega-i0,\mathbf{k})$. When the frequency dependent scattering rate
$\Gamma(\omega)$ is included, they acquire the form
\be \label{adv_retard_functions}
S^{(R,A)}(\omega,\mathbf{k})=e^{-\frac{c{\mathbf{k}}^{2}}{|eB|}}\sum\limits_{n=0}^\infty
(-1)^n\frac{S_n^{(R,A)}(\omega\pm
i\Gamma(\omega),\mathbf{k})}{(\omega\pm i\Gamma(\omega))^2-M_n^2}.
\ee
The scattering rate $\Gamma(\omega)$ is expressed via the retarded
fermion self-energy, $\Gamma(\omega)= -{\rm Im}\Sigma^R(\omega)$
which in general depends on the energy, temperature, field and the
Landau levels index $n$. This self-energy has to be determined
self-consistently from the Schwinger-Dyson equation. The exact form
of this equation actually depends on the model assumptions about the
impurity scattering, e.g. whether the impurity scatterers  are
short- or long-range and in many cases this equation is solved
numerically. Exactly this kind of consideration was made for
graphene in Ref.~\cite{Zheng2002PRB}, but in our paper we pursue
another goal which is to obtain a simple analytical expression for
the Hall conductivity. Accordingly here we chose a different
strategy. In Sec.~\ref{sec:conductivity-general} we derive general
expressions for both frequency dependent $\sigma_{xx}(\Omega)$ and
$\sigma_{xy}(\Omega)$ which include an unspecified frequency
dependent scattering rate $\Gamma(\omega)$. However, we make an
essential for the analytical work simplifying assumption that
$\Gamma(\omega)$ is independent of the Landau levels index $n$. This
assumption is justified when point-like impurity scattering is
considered \cite{Champel2002PRB}. The  expressions  obtained for
$\sigma_{xx}(\Omega)$ and $\sigma_{xy}(\Omega)$ are suitable for
investigation of microwave and optical conductivities in graphene.
Then in Sec.~\ref{sec:Hall} we consider the case of constant width
$\Gamma= \Gamma(\omega=0)= -{\rm Im}\Sigma^R(\omega=0) = 1/(2\tau)$,
where $\tau$ is the mean free time of quasiparticles and treat
$\Gamma$ as a phenomenological parameter. This approximation allows
one to obtain rather simple expressions for the Hall conductivity in
the various limits.

\section{General representation for electrical conductivity}
\label{sec:conductivity-general}

The frequency dependent electrical conductivity tensor is calculated
using the Kubo formula
\begin{equation}
\label{Kubo-cond} \sigma_{ij} (\Omega)= \frac{{\rm
Im}\Pi^R_{ij}(\Omega+i0)}{\Omega},
\end{equation}
where $\Pi^R_{ij}(\Omega)$ is the retarded current-current
correlation function obtained by analytical continuation
($\Pi^{R}_{ij}(\Omega )=\Pi_{ij}(i\Omega _{m}\to \Omega +i0)$) of
the imaginary time expression:
\begin{equation}
\label{im-time_Pi} \Pi_{ij}(i\Omega
_{m})=\frac{1}{V}\int\limits_{0}^{\beta }d\tau e^{i\Omega _{m}\tau
}\langle T_{\tau }J_{i}(\tau )J_{j}(0)\rangle ,\quad J_{i}(\tau
)=\int d^{2}r j^{i}(\tau,\mathbf{r} ), \quad \Omega_m= 2\pi m T.
\end{equation}
Here $V$ is the volume of the system, $\beta =1/T$ is the inverse
temperature, and $j^i(\tau,\mathbf{r})$ is the electric current
density operator
\begin{equation}
j^i(\tau,\mathbf{r}) = -\frac{\delta \mathcal{L}}{\delta A_i}=
-ev_{F}\sum_\sigma\bar{\Psi}_\sigma(\tau,\mathbf{r})\gamma^i\Psi_\sigma(\tau,\mathbf{r}).
\end{equation}
The brackets in Eq.~(\ref{im-time_Pi}) denote the averaging in the
grand canonical ensemble. Neglecting the impurity vertex
corrections, the calculation of the conductivity reduces to the
evaluation of the bubble diagram
\begin{equation}
\label{polar_oper} \Pi_{ij} (i\Omega _{m})=-e^2v_F^2T\sum_{n=-\infty
}^{\infty }\int \frac{d^{2}k}{(2\pi )^{2}}{\rm tr}\left[
\gamma^{i}S(i\omega_{n},{\mathbf{k}})\gamma^{j}S(i\omega_{n}-i\Omega_{m},{\mathbf{k}})\right],
\end{equation}
where $\mbox{tr}$ includes also the summation over flavor index and
$S(i\omega ,\mathbf{k})$ reads
\begin{equation}
\label{spectr_repr} S(i\omega
_{n},{\mathbf{k}})=\int\limits_{-\infty }^{\infty }\frac{d\omega
\,A(\omega ,\mathbf{k})}{i\omega _{n}+\mu-\omega },
\end{equation}
with the spectral function given by the discontinuity relation
\be \label{spectral-function}A(\omega,\mathbf{k})=\frac{1}{2\pi
i}\left[S^A(\omega,\mathbf{k}) -S^R(\omega,\mathbf{k})\right] \ee
for $S^{A,R}(\omega,\mathbf{k})$ defined in
Eq.~(\ref{adv_retard_functions}). Note that the translation
non-invariant phase of the fermion Green's function (\ref{Green})
cancels out in $\Pi$.

The sum over Matsubara frequencies in Eq.~(\ref{polar_oper}) is
easily evaluated  when the fermion Green's function is written
using the spectral representation (\ref{spectr_repr}).
After this is done the analytical continuation is easily performed
and we obtain
\begin{equation}
\label{electric_cond}
\Pi_{ij}(\Omega+i0)={e^2v_F^2}\int\limits_{-\infty}^\infty d\omega
d\omega^\prime\frac{n_F(\omega^\prime)-n_F(\omega)}
{\omega-\omega^\prime-\Omega-i0}\int\frac{d^2k}{(2\pi)^2}{\rm
tr}\left[\gamma^iA(\omega,\mathbf{k})\gamma^j
A(\omega^\prime,\mathbf{k})\right],
\end{equation}
where $n_F(\omega)$ is the Fermi distribution function
$n_F(\omega)= 1/(\exp((\omega-\mu)/T)+1)$. The representation
(\ref{electric_cond}) is  suitable for studying both diagonal (see
Refs.~\cite{Gorbar2002PRB,Ferrer2003EPJB,Sharapov2003PRB,Gusynin2005PRB})
and off-diagonal conductivities. In Appendix~\ref{sec:A} we
generalize the calculations of the previous papers and obtain both
diagonal ac conductivity,
\begin{equation}
\label{optical-diagonal}
\begin{split}
\sigma_{xx}(\Omega)&
=\frac{e^2N_f}{4\pi^2\,\Omega}\int\limits_{-\infty}^\infty d\omega
[n_F(\omega)-n_F(\omega^\prime)]{\rm Re}\left\{\frac{2B}{\Delta^2-
(\omega+i\Gamma)^2}\left[\Xi_{1}(-B)-\Xi_{2}(-B) \right] \right.\\
& \left. +\left[\Xi_{1}(-B)+\Xi_{1}(+B)-\Xi_{2}(-B)-\Xi_{2}(+B)
\right]\psi\left(\frac{\Delta^2-(\omega+i\Gamma)^2}{2B}\right)+(\omega\leftrightarrow
\omega^\prime, \Gamma\leftrightarrow\Gamma^\prime) \right\},
\end{split}
\end{equation}
and off-diagonal ac conductivity
\begin{equation}
\label{optical-Hall}
\begin{split}
& \sigma_{xy}(\Omega) =-\frac{e^2N_f{\rm
sgn}(eB)}{4\pi^2\,\Omega}\,{\rm Im}\int\limits_{-\infty}^\infty
d\omega\Biggl\{[n_F(\omega)-n_F(\omega^\prime)] \\
&\times \left[ \Xi_{2}(-B) \frac{2B}{\Delta^2- (\omega+i\Gamma)^2}
+(\Xi_{2}(-B) -
\Xi_{2}(B))\psi\left(\frac{\Delta^2-(\omega+i\Gamma)^2}{2B}\right)
+(\omega\leftrightarrow \omega^\prime,
\Gamma\leftrightarrow\Gamma^\prime) \right]\\
& + [n_F(\omega)+n_F(\omega^\prime)]\left[ \Xi_{1}(-B)
\frac{2B}{\Delta^2- (\omega+i\Gamma)^2} + (\Xi_{1}(-B) -
\Xi_{1}(B))\psi\left(\frac{\Delta^2-(\omega+i\Gamma)^2}{2B}\right)-(\omega\leftrightarrow
\omega^\prime, \Gamma\leftrightarrow\Gamma^\prime) \right]\Biggr\}.
\end{split}
\end{equation}
Here $\psi$ is the digamma function, and we denoted $B\equiv
v_F^2|eB|/c$, $\omega^\prime=\omega+\Omega$,
$\Gamma=\Gamma(\omega),\Gamma^\prime=\Gamma(\omega^\prime)$ and
introduced the following short-hand notations
\begin{equation}\label{Xi}
\begin{split}
\Xi_{1}(\pm B) \equiv
\Xi_{1}(\omega,\omega^\prime,\Gamma,\Gamma^\prime,\pm B)=
\frac{(\omega^\prime+i\Gamma^\prime)(\omega+i\Gamma)-\Delta^2}
{[\omega-\omega^\prime+i(\Gamma-\Gamma^\prime)][\omega+\omega^\prime+i(\Gamma+\Gamma^\prime)]\pm 2B},\\
\Xi_{2}(\pm B) \equiv
\Xi_{2}(\omega,\omega^\prime,\Gamma,\Gamma^\prime,\pm B)=
\frac{(\omega^\prime-i\Gamma^\prime)(\omega+i\Gamma)-\Delta^2}
{[\omega-\omega^\prime+i(\Gamma+\Gamma^\prime)][\omega+\omega^\prime+
i(\Gamma-\Gamma^\prime)] \pm 2B}.
\end{split}
\end{equation}
To familiarize oneself with Eqs.~(\ref{optical-diagonal}) and
(\ref{optical-Hall}) let us firstly consider the diagonal
conductivity in two simple cases.

\begin{figure}[h]
\centering{
\includegraphics[width=8cm]{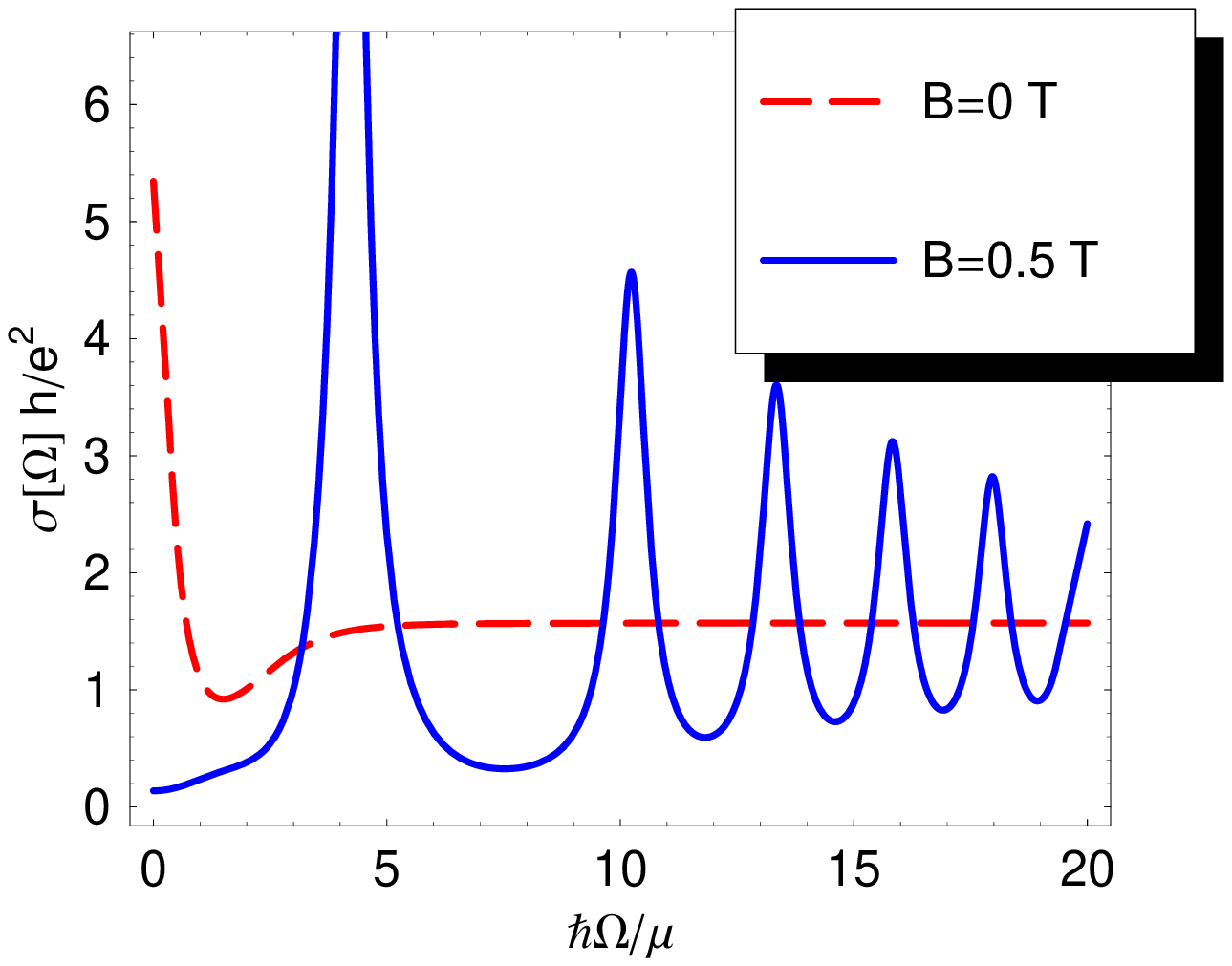}}
\caption{(Color online) The optical conductivity
$\sigma_{xx}(\Omega)$ measured in $e^2/h$ units as a function of
the frequency $\hbar \Omega/\mu$ for two different values of $B$
for $\mu = 50 \mbox{K}$, $T = 15\mbox{K}$ and $\Gamma = 10
\mbox{K}$. We use $eB \to (4.5\times10^4 \mbox{K}^2)
B(\mbox{T})$.} \label{fig:3-1}
\end{figure}

{\em Influence of the Landau quantization on the optical
conductivity.} We postpone a more comprehensive study of the
influence of the gap $\Delta$ and the form of the function
$\Gamma(\omega)$ on the ac conductivity for the future publication.
Here to illustrate the behavior of $\sigma_{xx}(\Omega)$ described
by Eq.~(\ref{optical-diagonal}) in Fig.~\ref{fig:3-1} we show the
results only for $\Delta=0$ and $\Gamma(\omega) = \mbox{const}$. One
can see that for $B=0$ there is Drude peak. However, when the
magnetic field is applied the spectral weight is transferred from
the Drude peak to the resonance peaks in the agreement with a recent
experiment \cite{Li2005}.

\subsection{dc limits of the longitudinal and Hall conductivities}
\label{sec:long-dc}

Let us firstly consider the dc limit of $\sigma_{xx}(\Omega)$,
\begin{equation}
\label{sigma_xx} \sigma_{xx}(\mu,B,T) = \sigma_{xx}(\Omega\to0) =
e^2 N_f \int \limits_{-\infty}^{\infty} d
\omega[-n_F^\prime(\omega)] \mathcal{A}_L(\omega,B,\Gamma,\Delta) ,
\end{equation}
where
\begin{equation}
\label{A}
\begin{split}
& \mathcal{A}_L(\omega,B,\Gamma,\Delta) =  \frac{1}{\pi^2}
\frac{\Gamma^2}{(v_F^2 e B/c)^2 + (2 \omega \Gamma)^2} \\
& \times \left\{ 2 \omega^2 + \frac{(\omega^2 + \Delta^2 +
\Gamma^2)(v_F^2 eB/c)^2 - 2 \omega^2 (\omega^2 - \Delta^2 +
\Gamma^2) v_F^2 |eB|/c}
{(\omega^2 - \Delta^2 - \Gamma^2)^2 + 4 \omega^2 \Gamma^2} \right. \\
&\left. - \frac{\omega(\omega^2 - \Delta^2 + \Gamma^2)}{\Gamma}
\mbox{Im} \psi \left( \frac{\Delta^2 - (\omega +  i \Gamma)^2}
{2v_F^2 |eB|/c}\right) \right\},
\end{split}
\end{equation}
and  $-n_F^\prime(\omega) =(1/4T)\cosh^{-2} [(\omega - \mu)/2T]$ is
the derivative of the Fermi distribution. Here the scattering rate
$\Gamma(\omega)$ remains a frequency dependent quantity. The
expression (\ref{sigma_xx}) was originally derived in
Refs.~\cite{Gorbar2002PRB,Sharapov2003PRB} (see also
\cite{Ferrer2003EPJB} for a related derivation of the thermal
conductivity) under the assumption $\Gamma(\omega) = \mbox{const}$.

Similarly to the dc expression (\ref{sigma_xx}) for $\sigma_{xx}$
one can take the dc limit $\Omega \to 0$ in Eq.~(\ref{optical-Hall})
and obtain
\be \label{sigma_xy-f}
\begin{split}
\sigma_{xy}(\mu,B,T)=& \frac{e^2N_f{\rm
sgn}(eB)}{2\pi^2}\int\limits_{-\infty}^\infty
d\omega\left\{\left(-n_F(\omega)\right)^\prime\frac{B(\Gamma^2-\Delta^2+\omega^2)}
{B^2+4\omega^2\Gamma^2}\left[\frac{2\omega\Gamma(B+\Gamma^2+\Delta^2-\omega^2)}
{[\Gamma^2+(\Delta-\omega)^2][\Gamma^2+(\Delta+\omega)^2]}
\right.\right.\\
&+\left.\left.{\rm
Im}\psi\left(\frac{\Delta^2-(\omega+i\Gamma)^2}{2B}\right)\right]
+n_F(\omega){\rm
Im}\left[(1+i\Gamma^\prime)\left(\frac{2(\omega+i\Gamma)}{B}
+\frac{2(\omega+i\Gamma)}{\Delta^2-(\omega+i\Gamma)^2}
\right.\right.\right.\\
&-\left.\left.\left.\frac{(\omega+i\Gamma)(\Delta^2-(\omega+i\Gamma)^2)}{{B^2}}
\psi^\prime
\left(\frac{\Delta^2-(\omega+i\Gamma)^2}{2B}\right)\right)\right]\right\},
\end{split}
\ee
where now $\Gamma^\prime\equiv d\Gamma(\omega)/d\omega$.  The term
with $n_F(\omega)$ can be integrated by parts and we finally arrive
at
\be \label{sigma_xy_final} \sigma_{xy}=e^2N_f{\rm
sgn}(eB)\int\limits_{-\infty}^\infty
d\omega[-n_F^\prime(\omega)]\mathcal{ A}_H(\omega,B,\Gamma,\Delta),
\ee
where
\be \label{AH}
\begin{split}
& \mathcal{A}_H(\omega,B,\Gamma,\Delta)= \frac{1}{2 \pi^2}
\left\{\frac{B(\Gamma^2-\Delta^2+\omega^2)}
{B^2+4\omega^2\Gamma^2}\frac{2\omega\Gamma(B+\Gamma^2+\Delta^2-\omega^2)}
{[\Gamma^2+(\Delta-\omega)^2][\Gamma^2+(\Delta+\omega)^2]}
+\frac{2\omega\Gamma}{B}+\arctan\frac{\Delta+\omega}{\Gamma}
-\arctan\frac{\Delta-\omega}{\Gamma}\right.\\
& -\left.2{\rm
Im}\ln\Gamma\left(\frac{\Delta^2-(\omega+i\Gamma)^2}{2B}\right)
+{\rm Im}\left[\left(\frac{B(\Gamma^2-\Delta^2+\omega^2)}
{B^2+4\omega^2\Gamma^2}+\frac{\Delta^2-(\omega+i\Gamma)^2}{B}\right)
\psi\left(\frac{\Delta^2-(\omega+i\Gamma)^2}{2B}\right)\right]\right\}.
\end{split}
\ee
Recall that here $B \equiv v_F^2|e B|/c$. Since we are considering
non-interacting quasiparticles, the temperature and $\mu$
dependences of the conductivities (\ref{sigma_xx}) and
(\ref{sigma_xy_final}) are only contained in the derivative of the
Fermi distribution. The spectral function (\ref{AH}) for the Hall
conductivity turns out to be more complicated than the corresponding
function (\ref{A}) for $\sigma_{xx}$, therefore it is useful to
consider simple limiting cases and to establish the correspondence
between our answer and the results obtained by previous authors.

\section{Hall conductivity}
\label{sec:Hall}

\subsection{Classical limit $|eB| v_F^2/c \ll \Gamma^2, \mu^2$. Drude-Zener formula}
\label{sec:Drude-Zener}

We begin our consideration with the classical limit $B \to 0$ (or
more exactly $|eB| v_F^2/c \ll \Gamma^2, \mu^2$), when Landau
quantization is not essential. Using the asymptotic expansions
\begin{equation}
\label{psi-asymp}
\begin{split}
\ln \Gamma(z) &= \left(z-\frac{1}{2}\right) \ln z -z +
\frac{1}{2}\ln (2\,\pi )  + \frac{1}{12\,z} - \frac{1}{360\,z^3} +
{O}\left(\frac{1}{z} \right)^4, \\
 \psi(z)&= \ln z  -
\frac{1}{2\,z} - \frac{1}{12\,z^2} + \frac{1}{120\,z^4}+
O\left(\frac{1}{z^5}\right),
\end{split}
\end{equation}
we obtain that for $\Delta=0$ and $B \to 0$
\begin{equation}\label{F.B=Delta=0}
\mathcal{A}_H(\omega,B, \Gamma, 0) = \frac{1}{2 \pi^2}\frac{B
\Gamma}{1+ B^2/(4 \omega^2 \Gamma^2)} \frac{(3 \Gamma^4 - 8 \Gamma^2
\omega^2 - 3 \omega^4) \Gamma \omega - 3 (\Gamma^2 +\omega^2)^3
\arctan \frac{\omega}{\Gamma}}{6 \Gamma^3 \omega^2 (\Gamma^2 +
\omega^2)^2}.
\end{equation}
Accordingly for the $T=0$ conductivity we find
\be \label{sigma12-weak-field}
\sigma_{xy}(\mu,B,0)=-\frac{e^2v_F^2N_feB\mu^2 \, \mbox{sgn} \mu
}{2\pi c} \frac{1}{(v_F^2eB/c)^2 +4\mu^2\Gamma^2}, \qquad \Gamma \ll
|\mu|, \ee
where we restored $e v_F^2/c$. The diagonal conductivity
(\ref{sigma_xx}) in the same limit reads
\be \label{diagonal-sigma-Drude} \sigma_{xx}=\frac{\sigma_0}{1 +
(\omega_c \tau)^2}, \qquad \sigma_0=\frac{e^2N_f|\mu|}{4\pi\Gamma},
\ee
where the mean-free time of quasiparticles $\tau=1/2\Gamma$ enters
instead of the transport time, $\tau_{tr}$ because we ignored the
vertex corrections, and we introduced the cyclotron frequency,
\be \omega_c=\frac{|eB|v_F^2}{c|\mu|} = \frac{|eB|}{c m_c}, \qquad
m_c = \frac{|\mu|}{v_F^2}.\ee
Here in the second equality $\omega_c$ is written in terms of an
fictitious "relativistic" mass, $m_c$ which plays the role of the
cyclotron mass in the Lifshits-Kosevich formula \cite{Geim-new}.
The definition of $m_c$ shows that the chemical potential $|\mu|$
in graphene acquires also the meaning of the cyclotron mass, so
that the latter is easily tunable by the gate voltage
\cite{Geim-new}.

Accordingly the Hall conductivity (\ref{sigma12-weak-field}) can
also be written in the Drude-Zener form
\be \label{Hall-sigma-Drude}\sigma_{xy}=-\frac{\omega_c\tau\sigma_0
\, \mbox{sgn} (eB) \, \mbox{sgn}\mu}{1+(\omega_c\tau)^2}, \ee
 This result agrees with
Eq.~(23) of Ref.~\cite{Zheng2002PRB} when we take
$\tau_{tr}=\tau=1/2\Gamma$ and $N_f=1$. As we will see later for
$N_f =2$ all our results are twice bigger than the corresponding
results of Refs.~\cite{Zheng2002PRB,Shon1998JPSJ}.

Finally we consider the relationship (see Appendix~\ref{sec:D})
\begin{equation}
\label{number} \rho = \frac{N_f}{2\pi \hbar^2 v_F^2} \mu^2 {\rm sgn
\mu}
\end{equation}
between the chemical potential $\mu$ and carrier imbalance
(density) $\rho$ for the relativistic quasiparticles [$\hbar$ is
restored]. This relationship is in agreement with Fig.~3~d of
Ref.~\cite{Geim-new} and with Ref.~\cite{Kim-new}, viz.
 the cyclotron mass
in graphene, $m_c = (\pi \hbar^2 |\rho|/ v_F^2)^{1/2}$ is indeed
$\sim \sqrt{|\rho|}$. Since  experiment \cite{Geim-new} shows also
that $\sigma_0 \sim |\rho|$ for $B=0$, one may conclude that the
carrier concentration dependence $\Gamma(\mu) \sim |\mu|^{-1}$ for
$\Gamma \ll |\mu|$.

For the Hall resistivity one obtains from
Eqs.~(\ref{diagonal-sigma-Drude}) and (\ref{Hall-sigma-Drude}) that
\be \label{classics}
   \rho_{xy}=-
   \frac{\sigma_{xy}}{\sigma_{xx}^2+\sigma_{xy}^2}=\frac{B}{ec\rho}.
\ee
Thus for weak magnetic field we arrive at the standard expression
for the Hall coefficient $R_H=1/ec\rho$ which does not depend on the
scattering mechanism and is used for measuring the carrier density.

\subsection{The limit $T \ll |\mu| \ll \sqrt{|eB| v_F^2/c},\Gamma$}

Another interesting and analytically treatable limit is $T \ll
|\mu|\ll \sqrt{|eB|},\Gamma$. Since for $T\to0$ the derivative
$-n_F^\prime(\omega)\to \delta(\omega-\mu)$, the Hall conductivity
$\sigma_{xy}$ is directly expressed via Eq.~(\ref{AH}) and we need
only
\begin{equation}
\label{AH_mu} \mathcal{A}_H(\mu,B, \Gamma, 0)\simeq \frac{\mu B
\Gamma}{2 \pi^2}\left(\frac{4(B+\Gamma^2)}{B^2\Gamma^2}-
\frac{2\Gamma^2}{B^3}\psi^\prime\left(\frac{\Gamma^2}{2B}\right)\right),
\qquad  |\mu|\ll \sqrt{B},\Gamma.
\end{equation}
Substituting in Eq.~(\ref{AH_mu}) the asymptotic expansion of
$\psi^\prime(z)$ obtained from Eq.~(\ref{psi-asymp}) in the limit
$\sqrt{|eB|} \ll \Gamma$ we have
\begin{equation}
\mathcal{A}_H(\mu,B, \Gamma, 0)\simeq -\frac{4\mu
B}{3\pi^2\Gamma^3}.
\end{equation}
Finally restoring the $ev_F^2/c$ factor, we obtain
\begin{equation}
\label{sigma12-mu-to-zero} \sigma_{xy}=-\frac{4e^2v_F^2N_f eB
\mu}{3\pi^2c\Gamma^3}, \qquad |\mu|\ll\sqrt{|eB| v_F^2/c}\ll
\Gamma.
\end{equation}
In the same limit the conductivity $\sigma_{xx}$ is ``universal''
\cite{Shon1998JPSJ,Gorbar2002PRB,foot3}
\be \label{sigma11-mu-to-zero} \sigma_{xx}=\frac{e^2N_f}{\pi^2}. \ee
Moreover, theoretically the universal value $(e^2/2\pi \hbar)
(4/\pi)$ (or $e^2/\pi h$ per each type of carriers) is expected even
in arbitrary fields \cite{Gorbar2002PRB}, because the $n=0$ Landau
level is field independent. (We note that $N_f=2$ when Zeeman
splitting is neglected and $N_f=1$ when it is taken into account).
The experiments \cite{Geim-new}, however, show a bigger value of the
conductivity per carrier type, $e^2/h$. Note also that in
Ref.~\cite{Suzuura2002} it is argued that for the long-range
impurities in graphene the weak-localization correction makes a
positive contribution to the conductivity $\sigma_{xx}$ that might
explain the mentioned difference.

In the opposite limit, $\sqrt{|eB|}\gg\Gamma$ we find from
Eq.~(\ref{AH_mu}) that ${\cal A}_H\simeq -2\mu/(\pi^2\Gamma)$ and,
accordingly,
\be \sigma_{xy}=-\frac{2 N_f e^2\mu}{\pi^2\Gamma}, \qquad |\mu|
\ll \Gamma \ll \sqrt{|eB| v_F^2/c}. \ee
This limit is important in the strong field (Hall) regime at small
carrier densities. It allows one to extract the impurities
scattering rate $\Gamma$ studying the dependence of the Hall
conductivity as a function of the chemical potential (or carrier
density) in the vicinity of the point where the gate voltage changes
its sign.

\subsection{Unusual quantization of the Hall conductivity in graphene}
\label{sec:PRL}

In this section we discuss a full derivation of Eqs.~(4) -- (6)
from Ref.~\cite{Gusynin2005PRL}. Analyzing Eq.~(6) of
Ref.~\cite{Gusynin2005PRL} we demonstrated  that the quantized
Hall conductivity in graphene is equal to odd multiples of $2
e^2/h$. Here we recapitulate the arguments of
Ref.~\cite{Gusynin2005PRL} discussing a few interesting moments
not mentioned there.

There are two ways to derive $\sigma_{xy}$ in the clean limit. The
first option is to take the limit $\Gamma \to 0$ directly in
Eq.~(\ref{electric_cond}) as was done in Ref.~\cite{Gorbar2002PRB}
and the second option is to use a general expression
(\ref{sigma_xy_final}). This option is considered in
Appendix~\ref{sec:B}, where we show that
\begin{equation}
\label{hall_cond} \sigma_{xy}=-\frac{e^2N_f{\rm
sgn}(eB)}{4\pi}\left[\tanh
\frac{\mu+\Delta}{2T}+\tanh\frac{\mu-\Delta}{2T}+2\sum\limits_{n=1}^\infty
\left(\tanh\frac{\mu+M_n}{2T}+\tanh\frac{\mu-M_n}{2T}\right)\right].
\end{equation}
Note that $\sigma_{xy}$ is an antisymmetric function of $\mu$.
Rearranging terms in Eq.~(\ref{hall_cond}) and using that
$\tanh(\omega-\mu)/2T =1-2 n_F(\omega)$ one can rewrite
Eq.~(\ref{hall_cond}) in terms of the Fermi distribution
\begin{equation}
\label{sigma-H-nf} \sigma_{xy}=-\frac{e^2N_f{\rm sgn}(eB)}{2\pi}
\sum\limits_{n=0}^\infty (2n+1)\left[ n_F(M_n)+n_F(-M_n)
-n_F(M_{n+1})- n_F(-M_{n+1})\right].
\end{equation}
This representation for $\sigma_{xy}$ is an equivalent of Eq.~(18)
from Ref.~\cite{Jonson1984PRB} derived for an ideal
two-dimensional electron gas
\begin{equation}
\label{sigma-Jonson} \sigma_{xy} = -\frac{e^2}{2 \pi}
\sum_{n=0}^\infty (n+1)[n_F(\omega_n^{\mathrm{nonrel}}) -
n_F(\omega_{n+1}^{\mathrm{nonrel}})], \qquad
\omega_n^{\mathrm{nonrel}} = \frac{e B}{mc}\left(n+\frac{1}{2}
\right),
\end{equation}
where $m$ is the effective mass of the carriers with a parabolic
dispersion law. The difference between the positions of Landau
levels and their degeneracy for the Dirac quasiparticles and for
nonrelativistic electron gas is encoded in the energies $M_n \sim
\sqrt{n}$ and $\omega_n^{\mathrm{nonrel}} \sim (n+1/2)$ with
$n=0,1,2,\ldots$, and in the different factors $2n+1$ and $n+1$ in
Eqs.~(\ref{sigma-H-nf}) and (\ref{sigma-Jonson}), respectively.

Now we rewrite  Eq.~(\ref{hall_cond}) as follows
\cite{Gorbar2002PRB}
\begin{equation}
\label{hall-clean-filling} \sigma_{xy}= -\frac{e^2 N_f
\mbox{sgn}(eB)\mbox{sgn} \mu}{2 \pi \hbar} \nu_B,
\end{equation}
with the filling factor  \cite{foot1}
\begin{equation}
\label{filling-quantum}
\mbox{sgn} \mu \, \nu_B  = \frac{1}{2}\left[\tanh
\frac{\mu+\Delta}{2T}+\tanh\frac{\mu-\Delta}{2T}
  +2\sum\limits_{n=1}^\infty
\left(\tanh\frac{\mu+M_n}{2T}+\tanh\frac{\mu-M_n}{2T}\right)\right].
\end{equation}
Since we are considering the quantized Hall conductivity we
restore Planck constant $h = 2\pi \hbar$ in
Eq.~(\ref{hall-clean-filling}) and in what follows. Taking for
definiteness $\mu
>0$, $\Delta=0$ and using that $\tanh(\omega/2T)={\rm
sgn}(\omega)$ for $T \to 0$, we obtain from
Eq.~(\ref{hall-clean-filling}) that
\begin{equation}\label{sigma-steps}
\sigma_{xy} =-\frac{e^2 N_f\mbox{sgn}(eB)}{2\pi\hbar} \left[1+ 2
\sum_{n=1}^{\infty} \theta(\mu -M_n) \right]=-\frac{e^2 N_f
\mbox{sgn}(eB)}{h}\left(1+2\left[\frac{\mu^2c} {2\hbar
|eB|v_F^2}\right]\right),
\end{equation}
where $[x]$ denotes the integer part of $x$. The usual
semi-phenomenological argumentation (see e.g.
Ref.~\cite{Hajdu.book}) for  the occurrence of the IQHE states
that in the presence of disorder when the dependence $\mu(\rho)$
becomes a smooth function, while the function $\sigma_{xy}(\mu)$
remains step-like. Accordingly, when in Eq.~(\ref{sigma-steps})
the spin degeneracy is counted by choosing $N_f=2$, we arrive at
the Hall quantization rule (\ref{Hall-Dirac}). The classical
(\ref{hall-clean}) and quantum (\ref{sigma-steps}) Hall
conductivities coincide only for the odd fillings, $\nu_B = 2n+1$
as shown in Fig.~\ref{fig:4-1}), where for comparison the
dependence of $\sigma_{xx}(\nu_B)$ is also plotted. As expected
the minima of $\sigma_{xx}$ also occur at $\nu_B = 2n+1$, while
the peaks of $\sigma_{xx}$ coincide with the steps of
$\sigma_{xy}(\nu_B)$.

The appearance  of the odd integer number in Eq.~(\ref{Hall-Dirac})
rather than simply integer fillings is caused by the fact that the
degeneracy of the $n=0$ Landau level is only half of the degeneracy
of the levels with $n>0$ (see Appendix~\ref{sec:E}). The lowest
Landau level in the irreducible representation of the $2+1$
dimensional Dirac theory is special, because depending on which of
two inequivalent irreducible representations is used it is occupied
either by the electrons (fermions) or holes (antifermions) while at
$n\ge1$ there are solutions of the Dirac equation describing  both
electrons and holes (see Eq.~(\ref{eq:5})). The Dirac Lagrangian
(\ref{Lagrangian}) which embeds a pair of independent $\mathbf{K}$
points on graphene's Fermi surface, is written using a parity
preserving $4\times4$ reducible representation of $\gamma$ matrices
that contains two irreducible representations with different
parities. Therefore, the Landau levels associated with these
$\mathbf{K}$ points merge into the full spectrum of the Dirac
fermions in graphene. The superposition of the two spectra
corresponding to the two inequivalent irreducible representations
clearly explains a halved degeneracy of the lowest Landau level in
graphene, because this level can be occupied by the holes from the
$\mathbf{K}$ point (when $\mu<0$) and electrons from the
$\mathbf{K}^\prime$ point (when $\mu>0$). This property of the $n=0$
level does not depend on whether the gap $\Delta$ has a finite value
or $\Delta =0$. On the other hand, the higher levels may contain
either electrons (when $\mu>0$) or holes (when $\mu<0$) both from
$\mathbf{K}$ and $\mathbf{K}^\prime$ points.

The other way to explain the origin of ``strange'' odd numbers is to
refer to, the mentioned above, positions of the minima of Shubnikov
de Haas oscillations of $\sigma_{xx}$. Their unusual positions are
caused by the phase shift of $\pi$ between the quantum magnetic
oscillations for the relativistic quasiparticles (see Refs.
\cite{Sharapov2004PRB,Gusynin2005PRB,Luk'yanchuk2004PRL}) and the
corresponding oscillations for the nonrelativistic quasiparticles.
The origin of the phase shift can be traced back to the different
quantization of the relativistic, $M_n \sim \sqrt{n}$ and
nonrelativistic $\omega_n^{\mathrm{nonrel}} \sim (n+1/2)$ Landau
levels \cite{Sharapov2004PRB}.

One can gain a deeper insight into this by considering the
operator $\mathbb{S}=(1/2) \int d^2 r[\Psi^\dagger(t,\mathbf{r}),
S_z \Psi(t,\mathbf{r})]$,  where $[,]$ is the commutator and the
matrix
\begin{equation}
S_z = I_2 \otimes (\sigma_3/2) = \left(
        \begin{array}{cc}
          \sigma_3/2 & 0 \\
          0 & \sigma_3/2 \\
        \end{array}
      \right).
\end{equation}
The operator $\mathbb{S}$ generates the rotations of spinors in the
plane by an angle $\phi$ that are described by the operator $U(\phi)
= \exp(i \phi \mathbb{S})$. It is natural to interpret $\mathbb{S}$
as the {\em pseudospin\/} operator, because we use the spinors that
are related to the presence of two sublattices in graphene. Each
$\mathbf{K}$ point of graphene's Fermi surface is characterized by a
two-component spinor and there are two inequivalent $\mathbf{K}$
points. There is a temptation to make a direct analogy between the
{\em pseudospin} operator $\mathbb{S}$ and the {\em spin} operator
in $3+1$ dimensional Dirac theory. This, however, is misleading
because the very notion of the spin in $2+1$ and $3+1$ dimensions is
meaningful only for a massive particle. For a massless particle in
$3+1$ dimension instead of spin one introduces the helicity which
characterizes the projection of its spin on the direction of
momentum. Since in $2+1$ dimensional case  one cannot make rotations
around the direction of the quasiparticle momentum lying in the
two-dimensional plane, the helicity concept for massless particles
is meaningless in this case. Indeed, one can check that the
pseudospin operator $\mathbb{S}$ is identically zero for free
massless Dirac particles.

Nevertheless, below we argue that for the massless quasiparticles
in graphene in an external magnetic field the pseudospin acquires
a new meaning closely related to the Berry's phase discussed in
the different context \cite{Luk'yanchuk2004PRL,Kim-new}.

One can obtain (see Appendix~\ref{sec:E}) for $eB>0$ and $N_f=1$
that
\begin{equation}
\label{pseudospin} \mathbb{S} =
\frac{1}{2}\left[\sum_{m=-\infty}^0\left(b_{0m}^\dagger b_{0m} -
c_{0m}^\dagger c_{0m}\right) + \sum_{n=1}^\infty
\sum_{m=-\infty}^n\frac{\Delta}{M_n}\left( a_{nm}^\dagger a_{nm}+
b_{nm}^\dagger b_{nm} - c_{nm}^\dagger c_{nm}  - d_{nm}^\dagger
d_{nm} \right)\right],
\end{equation}
where $a_{nm}$ ($b_{nm}$) are the annihilation operators of
fermions with energies $M_n$ given by Eq.~(\ref{Mn}) (antifermions
with energies $-M_n$) for $\mathbf{K}$ point and $c_{nm}$
($d_{nm}$) are annihilation operators of fermions (antifermions)
for $\mathbf{K}^\prime$ point. The quantum number $m \leq n$ in
Eq.~(\ref{pseudospin}) reflects the degeneracy of each level in
angular momentum. Interestingly in the limit $\Delta \to 0$ only
the $n=0$ level contributes to $\mathbb{S}$, so that the notion of
the pseudospin is meaningful only for the states from the lowest
Landau level, i.e. for the zero modes.

Let us now consider the rotation by the angle $\phi=2\pi$ of a
quasiparticle state $|n=0\rangle =b_{0m}^{\dagger}|0\rangle $ from
the $n=0$ level. Here $|0\rangle$ is the vacuum state. For
$\Delta=0$ one can show that $U(2\pi)|n=0\rangle = \exp(i \pi)
|n=0\rangle$, i.e. after the rotation by $2\pi$ the quasiparticle
state from the lowest Landau level changes its phase by $\pi$. On
the other hand, the states from the levels with $n \geq 1$ remain
invariant, because for $\Delta=0$ the operator $U(\phi)$ does not
contain the operators that can change these states. It occurs
exactly due to this Berry's phase shift for massless quasiparticles
from the $n=0$ level in the external field, that the carriers in
graphene cannot be considered as simply very light carriers with a
finite mass. Note that in $3+1$ dimensional Dirac theory all
fermionic states acquire the phase shift by $\pi$ after the rotation
by $\phi =2\pi$.

If the $n=0$ level had the same degeneracy as the higher levels, the
Hall conductivity would had been quantized in a more conventional
manner,
\begin{equation}
\label{Hall-semicond} \sigma_{xy}^{\mathrm{semicond}} = -
\frac{4e^2}{h}n, \qquad n =0,1,\ldots
\end{equation}
which one might expect for a two-band [the first band would
corresponds to the electrons with $\omega_n = M_n-\mu$ and the
second band, accordingly, to the holes with $\omega_n = - M_n-\mu$],
two-valley [corresponding to $\mathbf{K}$ and $\mathbf{K}^\prime$
points of graphen's Fermi surface] semiconductor. Again in
Eq.~(\ref{Hall-semicond}) we assumed that $e,B,\mu>0$

It is appropriate here to mention that although the conventional
quantized Hall conductance $\sigma_{xy} = - (e^2/h) n$ with
$n=0,1,\ldots$ is often derived by solving the nonrelativistic
Schr{\"o}dinger equation, it was shown
\cite{MacDonald1983PRB,Nieto1985AJP} that there are no relativistic
corrections to this expression and the same result remains valid for
a relativistic electron gas confined in the plane described by the
$(3+1)$ dimensional Dirac equation. The case  of graphene is
different, because the Dirac theory is used to describe an {\em
effective theory\/} of non-relativistic quasiparticles with a linear
dispersion. Although one usually associates the Dirac-like
description of graphene with a linear dispersion of quasiparticle
excitations, $E(\mathbf{k}) = -\mu \pm v_F |\mathbf{k}|$ for $B=0$,
the quantization (\ref{Hall-Dirac}) survives even when there is a
nonzero gap $\Delta$.

Another important feature of graphene is that its Dirac-like
description is based on the $4 \times 4$ reducible, parity
preserving representation of $\gamma$ matrices which allows one to
include two inequivalent $\mathbf{K}$ points of graphene's Fermi
surface. From a theoretical point of view one may also choose a
separate $\mathbf{K}$ point and consider the role of the parity
breaking terms (see e.g.
Refs.~\cite{Haldane1988PRL,Schakel1991PRD}). This approach is
closely related to early unsuccessful \cite{Abouelsaood1985PRL}
attempts to explain the IQHE using the chiral anomaly. However, in
the case of the parity anomaly the Hall effect occurs even in zero
magnetic field and in the absence of Landau levels
\cite{Haldane1988PRL}. On the contrary, the Dirac-like description
of graphene preserves parity, so that the Hall conductivity is {\em
always absent\/} in zero magnetic field.

\subsection{Illustrations of analytical results for the Hall
conductivity}

\begin{figure}[h]
\centering{
\includegraphics[width=8cm]{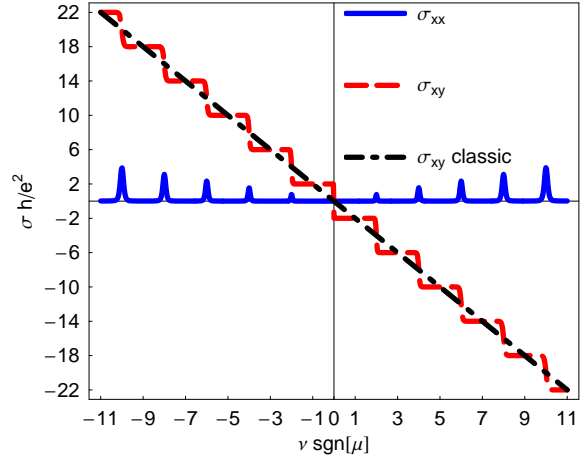}}
\caption{(Color online) The Hall  conductivity $\sigma_{xy}$ and
the diagonal conductivity $\sigma_{xx}$ measured in $e^2/h$ units
as a function of the filling $\nu_B$ for $T = 2\mbox{K}$, $\Gamma
= 1 \mbox{K}$ and $B = 2 \mbox{T}$. We use $eB \to (4.5\times10^4
\mbox{K}^2) B(\mbox{T})$ and assume that $\Delta=0$. The straight
line corresponds to a classical dependence $\sigma_{xy}= - ec
|\rho|\mbox{sgn} \mu/B $ [see Eq.~(\ref{hall-clean})]. }
\label{fig:4-1}
\end{figure}

Fig.~\ref{fig:4-1} is plotted to illustrate the  odd integer Hall
quantization (\ref{Hall-Dirac}) and it is computed on the base of
Eqs.~(\ref{sigma_xy_final}) and (\ref{AH}). Since the representation
(\ref{sigma_xy_final}), (\ref{AH})  is derived under the
approximations discussed in Secs.~\ref{sec:model} and
\ref{sec:conductivity-general}, this consideration ignores the
presence of localized states. Thus for finite scattering rate
$\Gamma$ this representation for $\sigma_{xy}(B,\Gamma,\Delta)$
provides only an approximate description of the Hall quantization,
because $\Gamma$ remains nonzero even between Landau levels.
Nevertheless, considering the relative simplicity and analytical
character of the two expressions (\ref{sigma_xx}) for $\sigma_{xx}$
and (\ref{sigma_xy_final}) for $\sigma_{xy}$, overall they give an
amazingly good description of the conductivities. In particular, one
can see $\sigma_{xx}$ is indeed very small in the plateaux regions
of $\sigma_{xy}$.

\begin{figure}[h]
\centering{
\includegraphics[width=8cm]{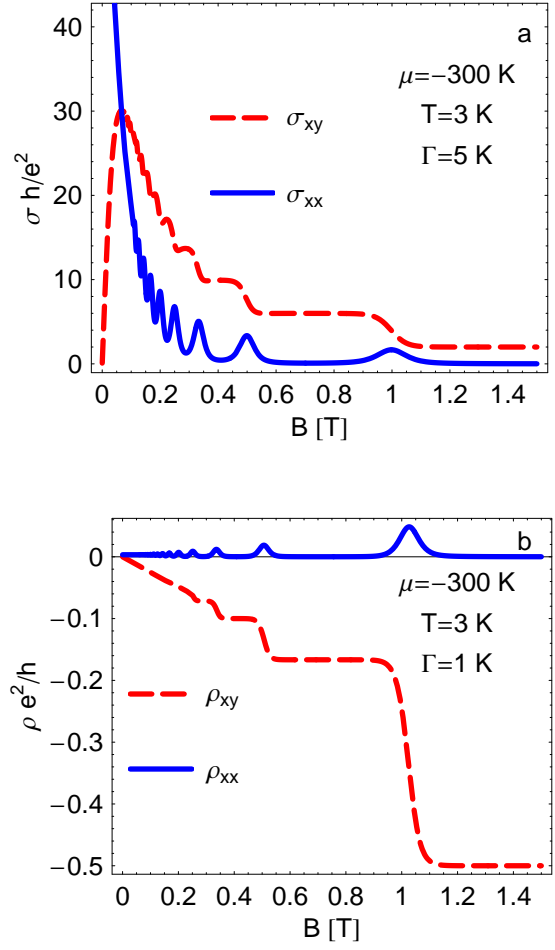}}
\caption{(Color online) (a) The Hall conductivity $\sigma_{xy}$
and the diagonal conductivity $\sigma_{xx}$ measured in $e^2/h$
units as a function of field $B$ for $\Gamma= 5 \mbox{K}$ for $\mu
= -300\mbox{K}$ and $T = 3 \mbox{K}$. (b) The Hall resistivity
$\rho_{xy}=-\sigma_{xy}/(\sigma_{xx}^2+ \sigma_{xy}^2)$ and the
diagonal resistivity $\rho_{xx}=\sigma_{xx}/(\sigma_{xx}^2+
\sigma_{xy}^2)$ measured in $h/e^2$ units as a function of field
$B$ for $\Gamma= 1 \mbox{K}$ for $\mu = -300\mbox{K}$ and $T = 3
\mbox{K}$. We use $eB \to (4.5\times10^4 \mbox{K}^2) B(\mbox{T})$
and assume that $\Delta=0$. } \label{fig:4-2}
\end{figure}

The IQHE can be obtained by varying either carrier concentration as
done in Fig.~\ref{fig:4-1} or the value of the applied magnetic
field. The latter possibility is shown in Fig.~\ref{fig:4-2}, where
both conductivities (Fig.~\ref{fig:4-2} a) and resistivities
(Fig.~\ref{fig:4-2} b) are plotted at $T = 3 \mbox{K}$. To observe
the quantum Hall effect in conventional semiconductors one should go
down to the temperatures lower than $10 \mbox{K}$, while in graphene
it can be observed up to $100 \mbox{K}$. This wider range of
temperatures is related to the fact that for the Dirac
quasiparticles the distance between Landau levels is much larger
than for the nonrelativistic quasiparticles in the same applied
field, so that some signatures of Shubnikov de Haas oscillations
remain notable even at room temperature \cite{Geim-new}.

Although the definition of the filling factor \cite{foot1} does not
depend on whether relativistic or nonrelativistic quasiparticles are
considered, the relationship (\ref{number}) between the chemical
potential and carrier imbalance shows that the small filling factors
become accessible in  relatively small, compared to conventional
semiconductors, fields. These features make the IQHE in graphene
very promising for fundamental research and possible applications.

Finally one can observe that $\sigma_{xy} =0$ for $B \to 0$. This
illustrates the point mentioned in Sec.~\ref{sec:PRL} that the IQHE
in graphene is conventional in the sense that its explanation does
not rely on any kind of parity breaking anomaly
\cite{Haldane1988PRL,Schakel1991PRD}.

\subsection{Magnetic catalysis and its observation in the Hall conductivity}
\label{sec:catalysis}

\begin{figure}[h]
\centering{
\includegraphics[width=8cm]{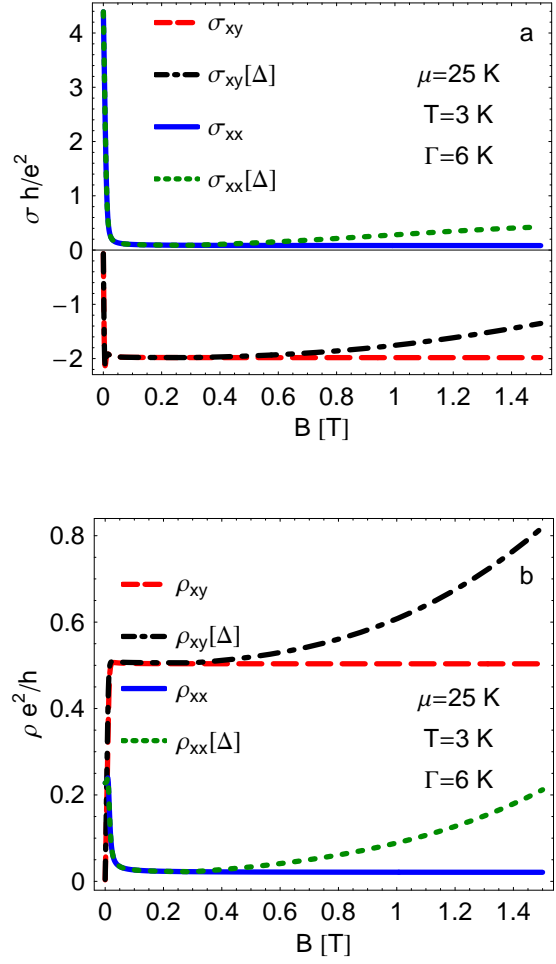}}
\caption{(Color online) The Hall conductivity $\sigma_{xy}$ and the
diagonal conductivity $\sigma_{xx}$ measured in $e^2/h$ units as a
function of field $B$ for $\Gamma= 6 \mbox{K}$ for $\mu =
25\mbox{K}$ and $T = 3 \mbox{K}$. The dash-dotted (black) and dotted
(green) lines are calculated using $\Delta(B)$ given by
Eq.~(\ref{gap-catalysis}). (b) The Hall resistivity $\rho_{xy}$ and
the diagonal resistivity $\rho_{xx}$ measured in $h/e^2$ units as a
function of field $B$ for $\Gamma= 6 \mbox{K}$ for $\mu =
25\mbox{K}$ and $T = 3 \mbox{K}$. The dash-dotted (black) and dotted
(green) lines are calculated using $\Delta(B)$ given by
Eq.~(\ref{gap-catalysis}). We use $eB \to (4.5\times10^4 \mbox{K}^2)
B(\mbox{T})$.} \label{fig:4-3}
\end{figure}

The flat, field independent behavior of $\sigma_{xx}(B)$ and
$\sigma_{xy}(B)$ for $B \gtrsim 1.1 \mbox{T}$ seen in
Fig.~\ref{fig:4-2} corresponds to the regime where only the lowest
Landau level is filled, because it always stays below the chemical
potential (except for $\mu =0$). As we already mentioned in
Sec.~\ref{sec:model}, it was predicted in
Refs.~\cite{Gusynin1995PRD,Gorbar2002PRB} that for Dirac fermions in
$2+1$ there is a phenomenon called {\em magnetic catalysis}. It is
expected that above a critical field $B_c$ which is a function of
$\mu$ and $T$, a gap $\Delta$ should open in the spectrum of the
Dirac fermions. Since the conditions for the magnetic catalysis  are
the most favorable at low carrier concentrations (or $\mu \approx
0$), in Fig.~\ref{fig:4-3} we present the behavior of
$\sigma_{xy}(B,\Delta(B))$ and $\sigma_{xx}(B,\Delta(B))$ for
$|\mu|$ smaller than was used to plot Fig.~\ref{fig:4-2}. For
comparison we plot these conductivities both for $\Delta=0$ and for
the phenomenological gap dependence
\be \label{gap-catalysis} \Delta(B) = c \sqrt{B - B_c}
\theta(B-B_c),\ee
where $c$  is some constant. As stated above, $B_c = B_c(\mu,T)$,
but for illustrative purposes we simply choose an arbitrary value
$B_c = 0.3 \mbox{T}$. As one can clearly see, the opening of the gap
$\Delta$ causes the decrease of $|\sigma_{xy} (B)|$ from the last
plateau value $2 e^2/h$. This tendency can also be understood from
Eq.~(\ref{hall_cond}). In the strong field limit the sum over $M_n$
with $n>0$ does not contribute and we obtain (for $eB>0$ and
$\mu>0$)
\begin{equation}
\label{catalysis} \sigma_{xy} = - \frac{e^2 N_f}{4 \pi
\hbar}\left[\tanh \frac{\mu+\Delta(B)}{2T} + \tanh \frac{\mu -
\Delta(B)}{2T}\right] \to - \frac{e^2 N_f}{2 \pi \hbar} \theta(\mu -
\Delta(B)), \qquad T \to 0.
\end{equation}
Eq.~(\ref{catalysis}) shows that when $\Delta(B)>\mu$ the magnetic
catalysis leads to a formation of a new insulating phase. On the
other hand,  we observe in Fig.~\ref{fig:4-3}~(a) that
$\sigma_{xx}(B)$ increases as the gap opens, while the increase of
the diagonal resistivity (Fig.~\ref{fig:4-3}~(b)) also indicates the
system goes towards an insulating phase.

In our consideration we assumed that the opening of the gap $\Delta$
does not affect the chemical potential $\mu$. We will come back to
this important issue in Sec.~\ref{sec:detection}.

\section{Hall angle and Nernst coefficient}
\label{sec:Nernst}

The  approach presented  allows one to calculate other transport
coefficients such as thermal conductivity
$\kappa_{ij}(B,\Gamma,\Delta)$ [see
Refs.~\cite{Ferrer2003EPJB,Sharapov2003PRB,Gusynin2005PRB}, where
the diagonal thermal conductivity is studied] and Peltier
(thermoelectric) conductivity $\beta_{ij} (B,\Gamma,\Delta)$
tensors. The calculation of the off-diagonal coefficients involves
some subtleties, because the conventional Kubo expressions have to
be altered \cite{Smrcka1997JPC,Oji1985PRB,Cooper1997PRB} to reflect
the role of magnetization on the electronic thermal transport in
applied field. However in the low-temperature limit one may rely on
the Sommerfeld expansion and express the thermoelectric tensor
through the conductivity tensor
\begin{equation}
\label{thermoelectric-low-T} \beta_{ij} = - \frac{\pi^2}{3}
\frac{T}{e} \frac{\partial \sigma_{ij}}{\partial \mu}.
\end{equation}

The Nernst signal measured in the absence of electric current is
expressed in terms of $\sigma_{ij}$ and $\beta_{ij}$ as
\begin{equation}
\label{Nernst-voltage-def} e_y(T,B) \equiv - \frac{E_y}{\nabla_x T}
= \frac{\sigma_{xx} \beta_{xy} - \sigma_{xy}
  \beta_{xx}}{\sigma_{xx}^2 + \sigma_{xy}^2}.
\end{equation}
Since in the low-temperature limit Eq.~(\ref{thermoelectric-low-T})
is valid, the Nernst signal (\ref{Nernst-voltage-def}) can be found
differentiating the Hall angle
\begin{equation}\label{Hall-angle}
\Theta_H = \arctan \frac{\sigma_{xy}}{\sigma_{xx}}
\end{equation}
via the relation
\begin{equation}
\label{Hall-angle2Nernst} e_y(T,B) = - \frac{\pi^2}{3} \frac{T}{e}
\frac{\partial \Theta_H}{\partial \mu}.
\end{equation}
Here based on the results derived in the previous sections, we will
study the behavior of the Hall angle and Nernst coefficient.

\subsection{Drude-Zener $|eB| v_F^2/c \ll \Gamma^2 \ll \mu^2$ and
$T \ll |\mu| \ll \sqrt{|eB| v_F^2/c} \ll \Gamma$ limits}
\label{sec:Drude-Zener_Nernst}

In the classical limit (see Sec.~\ref{sec:Drude-Zener})
$\sigma_{xy}$ and $\sigma_{xx}$ are given by
Eqs.~(\ref{Hall-sigma-Drude}) and (\ref{diagonal-sigma-Drude}),
respectively. Hence,
\begin{equation}
\label{Hall-angle_Drude} \Theta_H \simeq \tan \Theta_H = -\omega_c
\tau \, \mbox{sgn}(eB)\,\mbox{sgn} \mu  =
-\frac{v_F^2eB}{2c\Gamma\mu}.
\end{equation}
The Hall angle (\ref{Hall-sigma-Drude}) diverges at $\mu \to 0$, but
one should consider only finite values of $\mu$, because it is
derived under the assumption $|eB| v_F^2/c \ll \Gamma^2 \ll \mu^2$.
Accordingly when $\mu$ is small, Eq.~(\ref{Hall-sigma-Drude}) is
valid only for very low fields $B$. When the character of the
carriers changes from hole-like ($\mu<0$) to electron-like
($\mu>0$), the Hall angle also changes from positive to negative.
The behavior of the Nernst signal
\be \label{Nernst_Drude}e_y=-\frac{\pi^2}{3}\frac{k_B^2Tv_F^2
B}{2c\Gamma\mu^2} \ee
mirrors the divergence of the Hall angle at $\mu\to0$, but $e_y <0$
irrespective of the sign of $\mu$. [Boltzmann constant $k_B$ is
restored in Eq.~(\ref{Nernst_Drude}).] Eqs.~(\ref{Hall-angle_Drude})
and (\ref{Nernst_Drude}) agree with the results obtained in
Ref.~\cite{Oganesyan2004PRB} using  Boltzmann theory. This is not
surprising, because this is exactly the limit described by
Drude-Zener theory. We note that although in
Ref.~\cite{Oganesyan2004PRB} the so called {\em $d$-density-wave}
state is considered, a direct comparison with our case is possible,
because the effective low-energy Dirac theory turns out to be the
same \cite{Sharapov2003PRB}.

Since in the regime $T \ll |\mu| \ll \sqrt{|eB| v_F^2/c} \ll
\Gamma$ the conductivities $\sigma_{xx}$ and $\sigma_{xy}$ are
given by Eqs.~(\ref{sigma11-mu-to-zero}) and
(\ref{sigma12-mu-to-zero}), respectively, for the Hall angle one
obtains
\be \label{angle-mu-to-zero} \Theta_H=
-\frac{4v_F^2eB\mu}{3c\Gamma^3}. \ee
Hence the Nernst signal is positive and given by
\be \label{nerst-mu-to-zero} e_y \simeq
\frac{4\pi^2}{9}\frac{k_B^2Tv_F^2B}{c\Gamma^3}. \ee
The Nernst signal (\ref{nerst-mu-to-zero}) can be very large in
clean system because $e_y \sim\tau^3$ \cite{Oganesyan2004PRB}.
This regime is accessible due to the fact that $\mu \approx 0$,
i.e. one may consider the large and positive Nernst signal as
another fingerprint of the Dirac quasiparticles. We mention that
since the Dirac quasiparticles emerge in the scenarios with
unconventional charge density waves (UCDW)
\cite{Nersesyan1989JLTP,Dora2003PRB}, in Ref.~\cite{Dora2003PRB}
the large and positive $e_y$ is regarded as a hallmark of UCDW.

\subsection{Illustrations of analytical results and detection of the gap $\Delta$ from the Hall
angle measurements} \label{sec:detection}

\begin{figure}[h]
\centering{
\includegraphics[width=8cm]{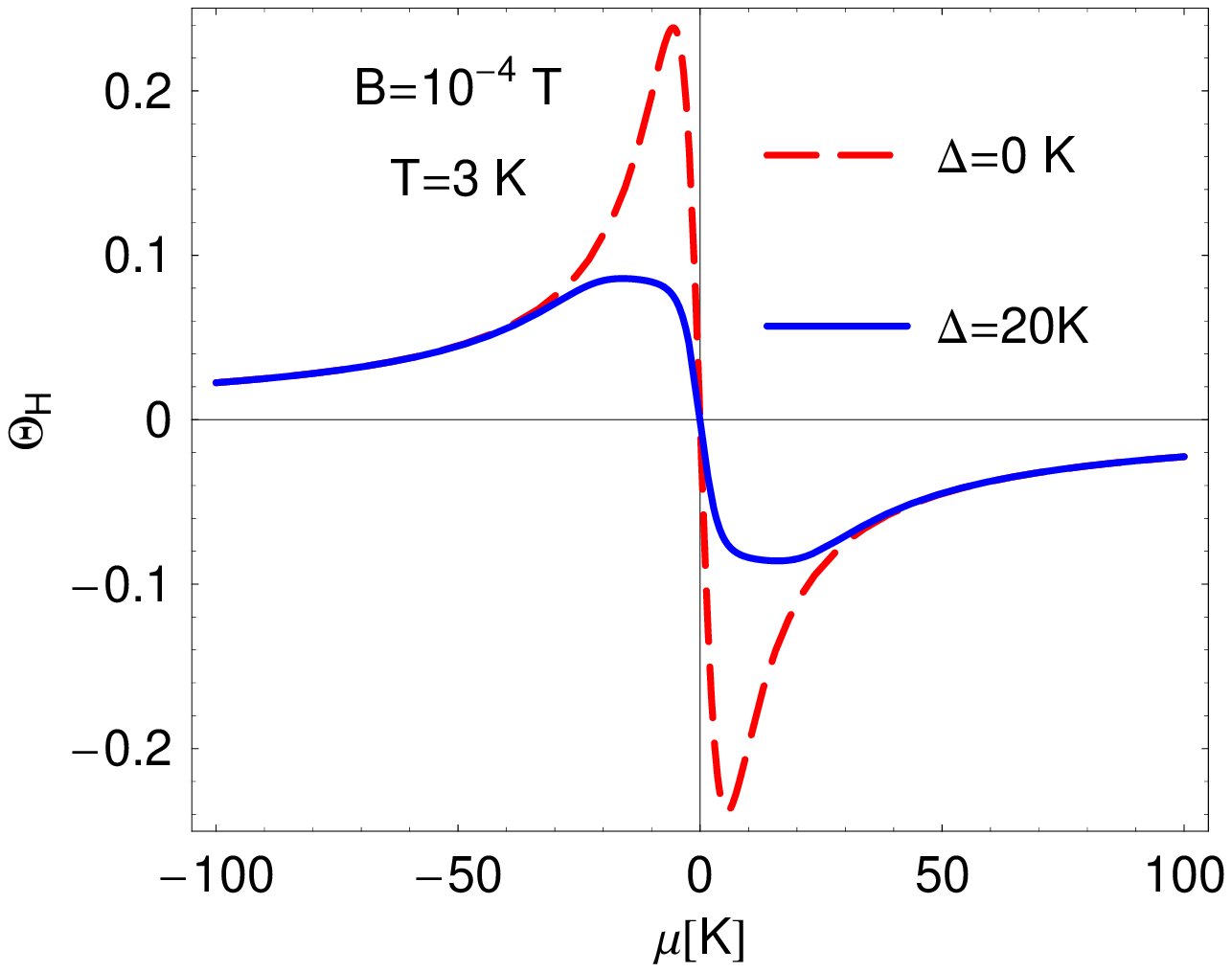}}
\caption{(Color online) The Hall angle $\Theta_H$
 as a function of chemical potential
$\mu$ for two different values of $\Delta$ for $B= 10^{-4}
\mbox{T}$, $T= 3 \mbox{K}$ and $\Gamma=1 \mbox{K}$. We use $eB \to
(4.5\times10^4 \mbox{K}^2) B(\mbox{T})$. } \label{fig:5-1}
\end{figure}

\begin{figure}[h]
\centering{
\includegraphics[width=8cm]{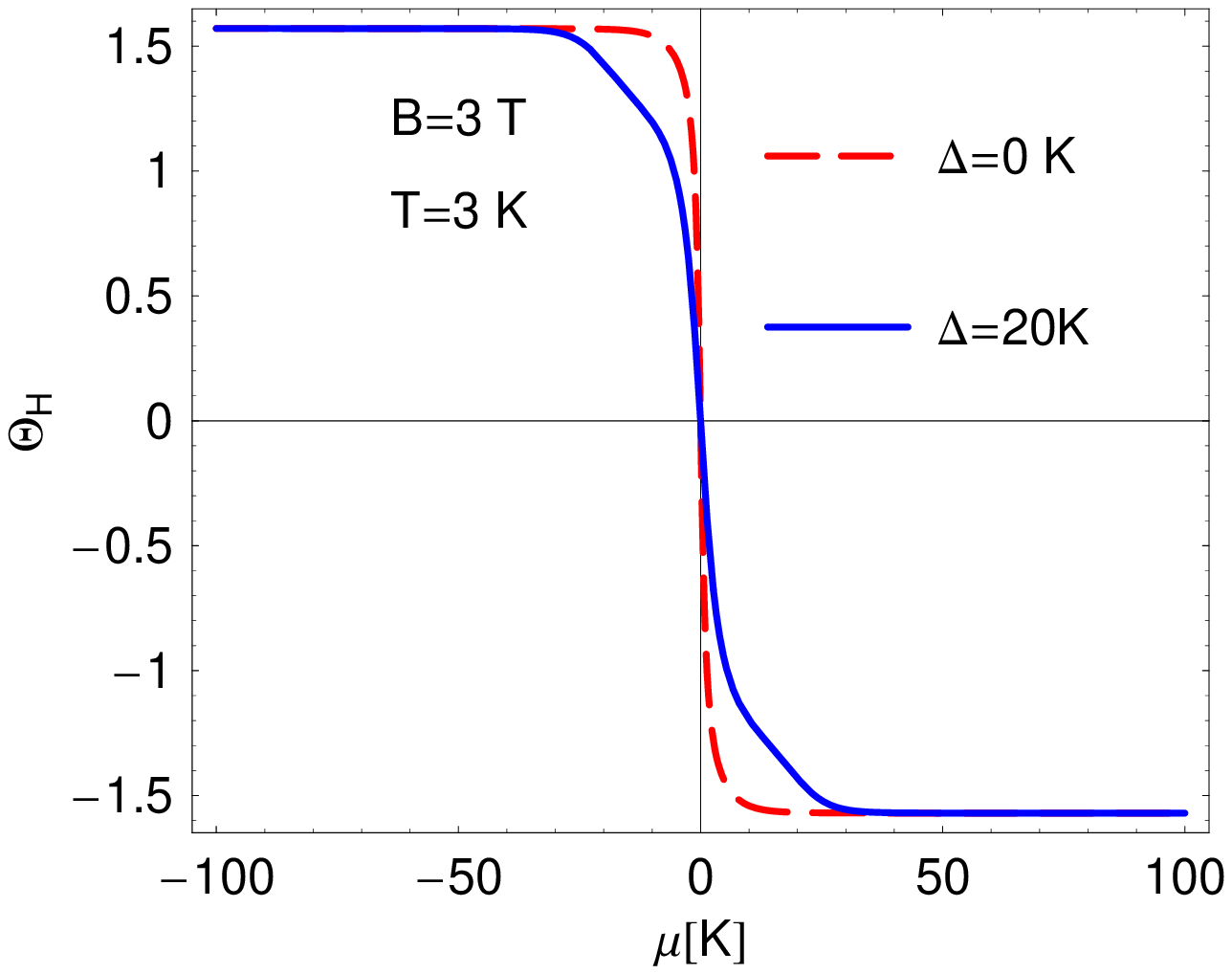}}
\caption{(Color online) The Hall angle $\Theta_H$
 as a function of chemical potential
$\mu$ for two different values of $\Delta$ for $B= 3 \mbox{T}$,
$T= 3 \mbox{K}$ and $\Gamma=1 \mbox{K}$. We use $eB \to
(4.5\times10^4 \mbox{K}^2) B(\mbox{T})$. } \label{fig:5-2}
\end{figure}

In Fig.~\ref{fig:5-1} and \ref{fig:5-2} we present the dependence
of the Hall angle (\ref{Hall-angle}) on $\mu$ for two different
values of the field $B$. The case of small $B$ shown in
Fig.~\ref{fig:5-1} agrees with the analytical expressions
discussed above. Indeed when $|\mu|$ decreases, there is an
increase of $\Theta_H$ [cf. Eq.(\ref{Hall-angle_Drude})] followed
by the regime where $\Theta_H$ crosses zero [cf.
Eq.~(\ref{angle-mu-to-zero})].

\begin{figure}[h]
\centering{
\includegraphics[width=8cm]{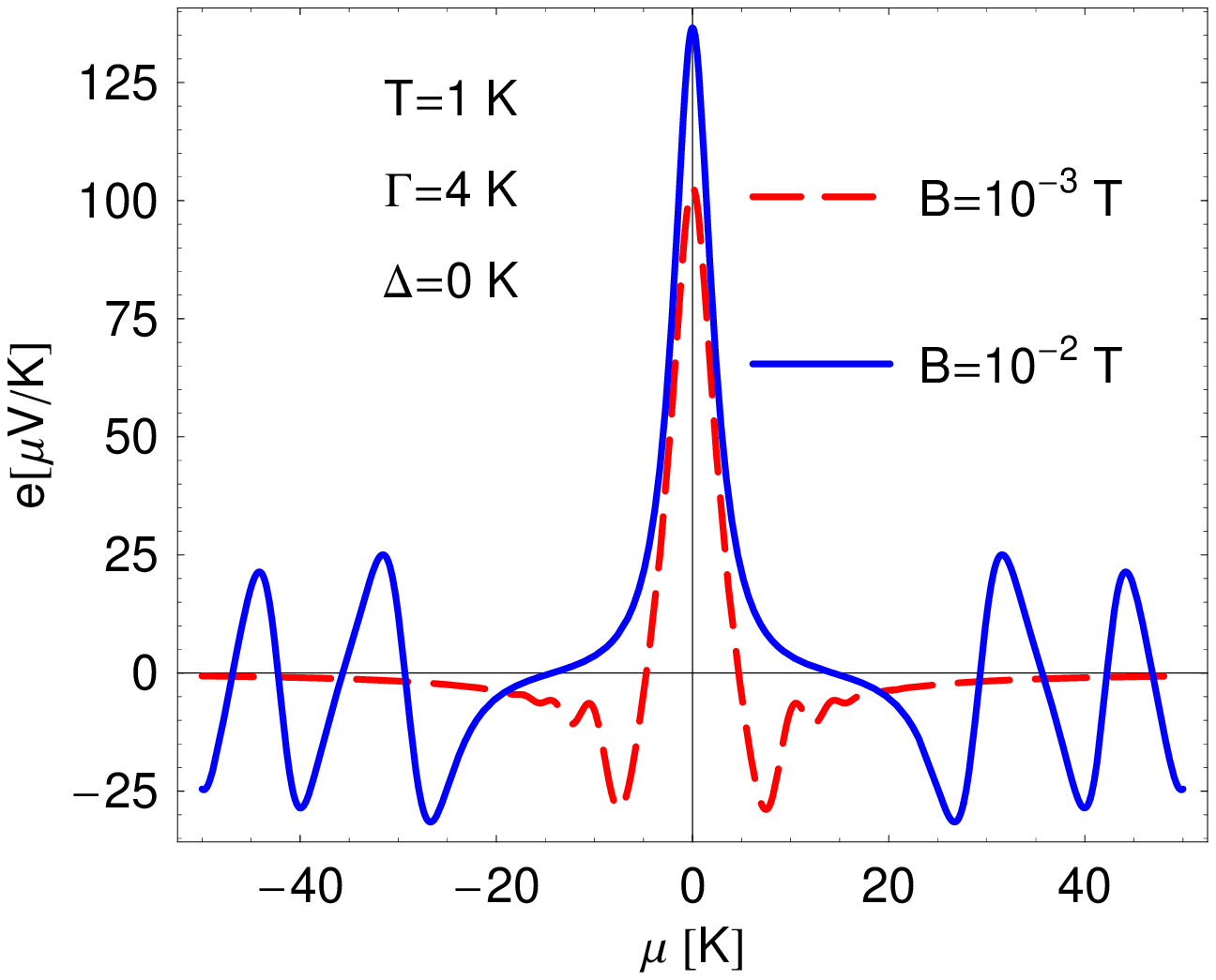}}
\caption{(Color online) The Nernst signal $e_y$ in $\mu V/K$
 as a function of chemical potential
$\mu$ for two different values of $B$ for $T= 1 \mbox{K}$,
$\Gamma=4 \mbox{K}$ and $\Delta =0 \mbox{K}$. We use $eB \to
(4.5\times10^4 \mbox{K}^2) B(\mbox{T})$. } \label{fig:5-3}
\end{figure}

\begin{figure}[h]
\centering{
\includegraphics[width=8cm]{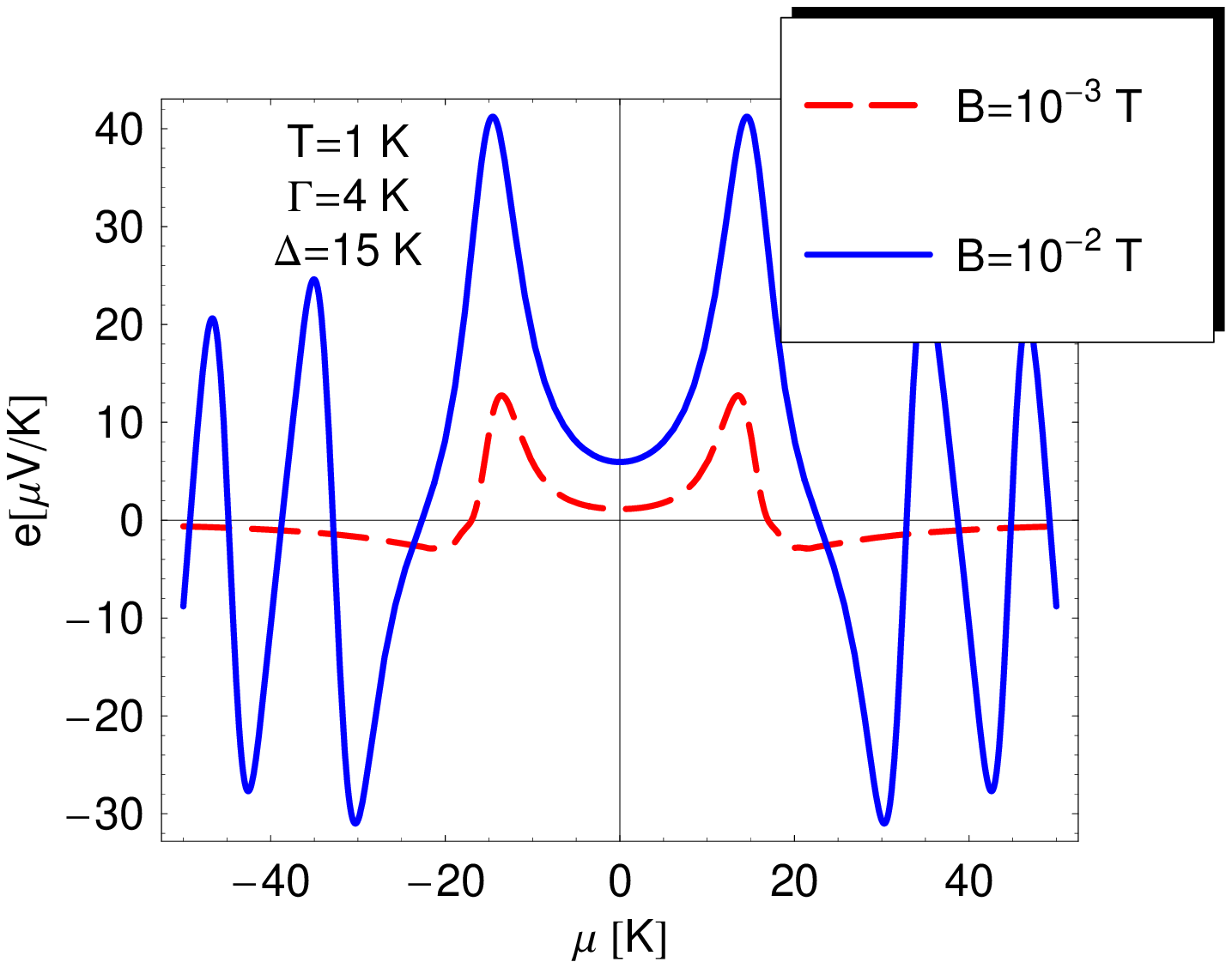}}
\caption{(Color online) The Nernst signal $e_y$ in $\mu V/K$
 as a function of chemical potential
$\mu$ for two different values of $B$ for $T= 1 \mbox{K}$,
$\Gamma=4 \mbox{K}$ and $\Delta =15 \mbox{K}$. We use $eB \to
(4.5\times10^4 \mbox{K}^2) B(\mbox{T})$. } \label{fig:5-4}
\end{figure}

In Fig.~\ref{fig:5-3} and \ref{fig:5-4} we show the behavior of
the Nernst signal. When $|\mu|$ is large and $B$ is small, $e_y<0$
and rather small. However, when $|\mu|$ decreases, $e_y$ becomes
positive and large ($\sim 100 \mu V/K$) in accord with
Eq.~(\ref{nerst-mu-to-zero}). When the field $B$ increases, the
value $e_y(\mu=0)$ becomes even larger and for finite $\mu$ there
are oscillations of $e_y(\mu)$.

Analyzing Figs.~\ref{fig:5-1} -- \ref{fig:5-4} one may discover
another interesting property, viz. in the presence of a nonzero gap
$\Delta$ the dependence $\Theta_H(\mu, \Delta)$ in
Figs.~\ref{fig:5-1}, \ref{fig:5-2} near $\mu \approx 0$ is not so
steep as compared to the $\Delta=0$ case. Accordingly this feature
is reflected in the Nernst signal, so that a finite $\Delta$ also
shows up in the dependence $e_y(\mu,\Delta)$ as a dip. Thus we
suggest that careful study of the Hall angle $\Theta_H(\mu, \Delta)$
and Nernst signal $e_y(\mu,\Delta)$ may help to establish the
presence of a nonzero gap $\Delta$ in graphene.

It is important, however, to stress that the gap is detectable only
if its opening does not change the chemical potential $\mu$. The
situation is exactly the same as in Ref.~\cite{Sharapov2003PRB},
where we considered the possibility of detecting the gap using
precise measurements of the period of the quantum magnetic
oscillations (de Haas van Alphen or Shubnikov de Haas). In the clean
system with a fixed carrier density $\rho$, the chemical potential
$\mu$ is given by the number equation (\ref{mu-n:eq}). The opening
of the gap $\Delta$ results in the adjustment of $\mu$, so that the
gap cannot be detected neither from the period of the oscillations
\cite{Sharapov2003PRB} nor from the Hall angle. Nevertheless, the
observation of the quantum Hall effect \cite{Geim-new,Kim-new} shows
that there is localization and the chemical potential remains fixed
making the gap detection possible.

Finally we stress that here we neglected the dependence of
$\Gamma(\mu)$, while a simple argument given at the end of
Sec.~\ref{sec:Drude-Zener} shows that this carrier concentration
dependence is rather important.

\section{Conclusions}
\label{sec:concl}

In this paper we have studied the dc Hall conductivity, Hall angle
and Nernst signal in a planar system with relativistic  Dirac-like
spectrum of quasiparticle excitations. We also presented the results
for the diagonal optical conductivity in the external magnetic field
perpendicular to the plane.

Our main results can be summarized as follows.\\
\indent (1) We have obtained analytical expressions for the
diagonal optical conductivity $\sigma_{xx}(\Omega,\mu,B,T,\Delta)$
[Eq.~(\ref{optical-diagonal})] and the off-diagonal optical
conductivity $\sigma_{xy}(\Omega,\mu,B,T,\Delta)$
[Eq.~(\ref{optical-Hall})].

\indent (2) We have derived the analytical expression
(\ref{sigma_xy_final}) with the kernel (\ref{AH}) for the dc Hall
conductivity $\sigma_{xy}(\mu,B,T,\Delta)$ which includes an
arbitrary impurity scattering rate $\Gamma(\omega)$ that was assumed
to be independent of the Landau level index $n$.

\indent (3) We have shown that in the classical limit our
expression for dc Hall conductivity Eq.~(\ref{sigma_xy_final})
reduces to the conventional Drude-Zener formula
(\ref{Hall-sigma-Drude}).

\indent (4) The direct comparison of the expression
(\ref{sigma-Jonson}) derived in Ref.~\cite{Jonson1984PRB} for Hall
conductivity in a two-dimensional electron gas with the
corresponding representation (\ref{sigma-H-nf}) for graphene allows
one to understand the origin of the odd integer Hall quantization
Eq.~(\ref{Hall-Dirac}) in graphene in terms of the difference
between the energies and degeneracies of the Landau levels in these
systems.

\indent(5) In Sec.~\ref{sec:PRL} we presented the arguments (using
the second quantization formalism) that the nontrivial Berry's
phase in graphene is associated with the anomalous properties of
the zero modes or the quasiparticles from the lowest Landau level.

\indent (6) We have investigated the behavior of the Hall angle
and the Nernst signal showing that for $\mu \approx 0$ there is an
interesting regime [see Eq.~(\ref{nerst-mu-to-zero})] where the
Nernst signal is strong and positive.

\indent (7) On the basis of the results obtained, we have discussed
in Secs.~\ref{sec:catalysis} and \ref{sec:detection} the possibility
of detecting a gap $\Delta$ that may open in the spectrum of the
Dirac-like quasiparticle excitations of graphene due to a nontrivial
interaction between them.

All our results are derived for noninteracting quasiparticles
treating the impurity scattering rate $\Gamma(\omega)$ as a
phenomenological parameter and without considering the interaction
with impurities that would demand solving an equation for
$\Gamma(\omega,\mu)$ (see Ref.~\cite{Zheng2002PRB}). Accordingly we
did not consider the problems related to  localization and a full
explanation of the IQHE (see Refs.~\cite{Hajdu.book,Prange-book}).
These problems by themselves acquire a new depth and deserve a
separate study. For example, it is pointed out in
Ref.~\cite{Geim-new} that localization effects in graphene are
suppressed. Interestingly, it is shown in Ref.~\cite{Suzuura2002}
that for long-range impurities the weak-localization correction
makes a positive contribution to the conductivity $\sigma_{xx}$.
This {\em anti-localization} property is related in
Ref.~\cite{Suzuura2002} to the Berry's phase of the Dirac fermions.
Definitely, such effects  were not considered in the present work.
Nevertheless we hope that the approach presented  here allows one to
explain in the most transparent way the difference between the Dirac
quasiparticles in graphene and nonrelativistic quasiparticles in
conventional semiconductors when these systems are placed in a
magnetic field.

\section{Acknowledgments}
We thank P.~Esquinazi, A.~Geim, W.A.~de~Heer, P.~Kim and
Y.~Kopelevich for discussing with us their experimental results
and J.P.~Carbotte, E.V.~Gorbar, V.M.~Loktev, V.A.~Miransky and
E.A.~Pashitsky for illuminating discussions. The work of V.P.G.
was supported by the SCOPES-project IB7320-110848 of the Swiss
NSF. S.G.Sh. was supported by the Natural Science and Engineering
Research Council of Canada (NSERC) and by the Canadian Institute
for Advanced Research (CIAR).

\appendix
\section{Calculation of $\sigma_{ij}(\Omega)$}
\label{sec:A}

The most efficient way of calculating $\sigma_{ij}(\Omega)$ is to
use again the spectral representation Eq.~(\ref{spectr_repr}) (for
$\mu=0$). This allows one to eliminate one of the integrations
over frequency in Eq.~(\ref{electric_cond}), so that for the real
part of the conductivity (\ref{Kubo-cond}) we obtain
\begin{equation}
\begin{split}
\sigma_{ij}(\Omega)&=\frac{e^2v_F^2}{2\pi\Omega}\, {\rm
Re}\int\limits_{-\infty}^\infty d\omega\,
\int\frac{d^2k}{(2\pi)^2}{\rm tr}\left\{[n_F(\omega)-
n_F(\omega+\Omega)]\gamma^i S^R(\omega+\Omega,\mathbf{k})
\gamma^jS^A(\omega,\mathbf{k})
\right.\\
&-\left. n_F(\omega)\gamma^i
S^R(\omega+\Omega,\mathbf{k})\gamma^jS^R(\omega,\mathbf{k})
+n_F(\omega+\Omega)\gamma^iS^A(\omega+\Omega,\mathbf{k})\gamma^jS^A(\omega,\mathbf{k})\right\}.
\end{split}
\end{equation}
Using the expressions Eq.~(\ref{adv_retard_functions}) for the
advanced and retarded Green's functions we obtain
\be \label{sigma-S_n}
\begin{split}
& \sigma_{ij}(\Omega)=\frac{e^2v_F^2}{2\pi\Omega}\,\,{\rm Re}
\int\limits_{-\infty}^\infty d\omega\int\frac{d^2k}{(2\pi)^2}
e^{-\frac{c{\mathbf{k}}^{2}}{|eB|}}\sum\limits_{n,m=0}^\infty(-1)^{n+m}
{\rm tr}\Biggl\{[n_F(\omega)-n_F(\omega+\Omega)]\\
& \times\left.\frac{\gamma^iS^R_n(\omega+\Omega)
\gamma^jS^A_m(\omega)}{[(\omega+\Omega+i\Gamma(\omega+\Omega))^2-M_n^2][(\omega-i\Gamma(\omega))^2-M_m^2]}
-n_F(\omega)\frac{\gamma^iS^R_n(\omega+\Omega)\gamma^jS^R_m(\omega)}
{[(\omega+\Omega+i\Gamma(\omega+\Omega))^2-M_n^2][(\omega+i\Gamma(\omega))^2-M_m^2]}\right.\\
&+n_F(\omega+\Omega)\frac{\gamma^iS^A_n(\omega+\Omega)\gamma^jS^A_m(\omega)}
{[(\omega+\Omega-i\Gamma(\omega+\Omega))^2-M_n^2][(\omega-i\Gamma(\omega))^2-M_m^2]}
\Biggr\},
\end{split}
\ee
where $S_n^{(R,A)}$ are the numerators of
Eq.~(\ref{adv_retard_functions}) obtained from Eq.~(\ref{S_n}) for
$S_n(i \omega_n,\mathbf{k})$ via the rule $S_n^{(R,A)}(\omega,
\mathbf{k}) = S_n(i \omega_n \to \omega \pm i \Gamma(\omega))$.
This allows to include the frequency dependent impurity scattering
rate, $\Gamma(\omega)$. The traces in Eq.~(\ref{sigma-S_n}) are
easily evaluated
\be
\begin{split}
\label{trace-S_n} &{\rm
tr}\left[\gamma^iS_n^{(R,A)}(\omega^\prime)\gamma^jS_m^{(R,A)}(\omega)\right]=
-8N_f[(\omega^\prime\pm i\Gamma^\prime)(\omega\pm i\Gamma)-\Delta^2]
\left[\delta_{ij}\left(L_n(x)L_{m-1}(x)+L_{n-1}(x)L_m(x)\right)\right.\\
&\left.+i\epsilon_{ij}\,{\rm
sgn}(eB)\left(L_n(x)L_{m-1}(x)-L_{n-1}(x)L_m(x)\right)\right]
-64N_f(2k_ik_j-\delta_{ij}\mathbf{k}^2)L_{n-1}^1(x)L_{m-1}^1(x),
\end{split}
\ee
where $\epsilon_{ij}$ is antisymmetric tensor ($\epsilon_{12} =1$)
and the argument of the Laguerre polynomials is
$x=2c{\bf{k}}^2/|eB|$. Integrating over momenta we obtain
\be \label{sigma-delta}
\begin{split}
\sigma_{ij}(\Omega)&=\frac{e^2v_F^2|eB|N_f}{2\pi^2c\,\Omega}\,\,{\rm
Re}
\sum\limits_{n,m=0}^\infty(-1)^{n+m+1}\left[\delta_{ij}(\delta_{n,m-1}+\delta_{n-1,m})+
i\epsilon_{ij}\,{\rm sgn}(eB)(\delta_{n,m-1}-\delta_{n-1,m})\right]\\
&\times \int\limits_{-\infty}^\infty
d\omega\left[\frac{[n_F(\omega)-n_F(\omega^\prime)]
[(\omega^\prime+i\Gamma^\prime)(\omega-i\Gamma)-\Delta^2]}{[(\omega^\prime+i\Gamma^\prime)^2
-M_n^2][(\omega-i\Gamma)^2-M_m^2]}-\frac{n_F(\omega)
[(\omega^\prime+i\Gamma^\prime)(\omega+i\Gamma)-\Delta^2]}{[(\omega^\prime+i\Gamma^\prime)^2
-M_n^2][(\omega+i\Gamma)^2-M_m^2]}\right.\\
&+\left.\frac{n_F(\omega^\prime)
[(\omega^\prime-i\Gamma^\prime)(\omega-i\Gamma)-\Delta^2]}{[(\omega^\prime-i\Gamma^\prime)^2
-M_n^2][(\omega-i\Gamma)^2-M_m^2]}\right].
\end{split}
\ee
The Kronnecker's delta symbols appeared in Eq.~(\ref{sigma-delta})
are due to the the orthogonality relation for the Laguerre
polynomials,
\begin{equation}
\label{Laguerre-orthogonality} \int_{0}^{\infty} d x e^{-x}
x^\alpha L_m^\alpha(x) L_n^\alpha (x) =
\frac{\Gamma(n+\alpha+1)}{n!}\delta_{mn},
\end{equation}
show that only transitions between neighboring Landau levels
contribute in the conductivity. The last term of
Eq.~(\ref{trace-S_n}) vanished after the angular integration. The
real part of Eq.~(\ref{sigma-delta}) reads
\be \label{sigma-Pi12}
\begin{split}
\sigma_{ij}(\Omega)&=\frac{e^2v_F^2|eB|N_f}{2\pi^2c\,\Omega}\sum\limits_{n,m=0}^\infty
(-1)^{n+m+1}\int\limits_{-\infty}^\infty
d\omega\left\{\delta_{ij}(\delta_{n,m-1}+\delta_{n-1,m})
[n_F(\omega)-n_F(\omega^\prime)]{\rm
Re}[\Pi^1_{n,m}(\omega,\omega^\prime)-
\Pi^2_{n,m}(\omega,\omega^\prime)]\right.\\
&-\left. \epsilon_{ij}\,{\rm
sgn}(eB)(\delta_{n,m-1}-\delta_{n-1,m})[[n_F(\omega)-n_F(\omega^\prime)]
{\rm
Im}\Pi^1_{n,m}(\omega,\omega^\prime)-[n_F(\omega)+n_F(\omega^\prime)]
{\rm Im}\Pi^2_{n,m}(\omega,\omega^\prime)] \right\},
\end{split}
\ee
where we introduced
\be
\Pi^1_{n,m}(\omega,\omega^\prime)=\frac{(\omega^\prime+i\Gamma^\prime)(\omega-i\Gamma)
-\Delta^2}{[(\omega^\prime+i\Gamma^\prime)^2-M_n^2][(\omega-i\Gamma)^2-M_m^2]},\quad
\Pi^2_{n,m}(\omega,\omega^\prime)=\frac{(\omega^\prime+i\Gamma^\prime)(\omega+i\Gamma)
-\Delta^2}{[(\omega^\prime+i\Gamma^\prime)^2
-M_n^2][(\omega+i\Gamma)^2-M_m^2]}. \ee
When we derived  Eq.~(\ref{sigma-Pi12}) we used the fact that the
real part of $\Pi^{1,2}_{n,m}$ does not alter when the
simultaneous replacements $i \Gamma \to - i \Gamma$ and  $i
\Gamma^\prime \to - i \Gamma^\prime$ are made, while its imaginary
part reverses sign. These features of $\sigma_{ij}(\Omega)$ are
also used below when Eqs.~(\ref{optical-diagonal}) and
(\ref{optical-Hall}) are written in the symmetric form. The sum
over $m$ in Eq.~(\ref{sigma-Pi12}) is easily calculated using
Kronnecker delta's,
\be
\begin{split}
\sigma_{ij}(\Omega)&=\frac{e^2v_F^2|eB|N_f}{2\pi^2c\,\Omega}\sum\limits_{n=0}^\infty
\int\limits_{-\infty}^\infty
d\omega\left\{\delta_{ij}[n_F(\omega)-n_F(\omega^\prime)]{\rm Re}
[\Pi^1_{n,n+1}(\omega,\omega^\prime)+\Pi^1_{n+1,n}(\omega,\omega^\prime)-
\Pi^2_{n,n+1}(\omega,\omega^\prime)\right.\\
&-\left.\Pi^2_{n+1,n}(\omega,\omega^\prime)] -\epsilon_{ij}\,{\rm
sgn}(eB)\left[[n_F(\omega)-n_F(\omega^\prime)]{\rm Im}
[\Pi^1_{n,n+1}(\omega,\omega^\prime)-\Pi^1_{n+1,n}(\omega,\omega^\prime)]\right.\right.\\
&-\left.\left.[n_F(\omega)+n_F(\omega^\prime)]{\rm
Im}[\Pi^2_{n,n+1}(\omega,\omega^\prime)-\Pi^2_{n+1,n}
(\omega,\omega^\prime)]\right]\right\}.
\end{split}
\ee
Further summation over $n$ can be performed expanding $\Pi^{1,2}$ in
terms of the partial fractions,
\be
\begin{split}
\Pi^1_{n,m}(\omega,\omega^\prime)&=\frac{(\omega^\prime+i\Gamma^\prime)(\omega-i\Gamma)
-\Delta^2}{[\omega-\omega^\prime-i(\Gamma+\Gamma^\prime)][\omega+\omega^\prime-
i(\Gamma-\Gamma^\prime)]+2B(n-m)}\left[\frac{1}{(\omega^\prime+i\Gamma^\prime)^2-M_n^2}-
\frac{1}{(\omega-i\Gamma)^2-M_m^2}\right],\\
\Pi^2_{n,m}(\omega,\omega^\prime)&=\frac{(\omega^\prime+i\Gamma^\prime)(\omega+i\Gamma)
-\Delta^2}{[\omega-\omega^\prime+i({\Gamma-\Gamma^\prime})][\omega+\omega^\prime+
i({\Gamma+\Gamma^\prime})]+2B(n-m)}\left[\frac{1}{(\omega^\prime+i\Gamma^\prime)^2-M_n^2}-
\frac{1}{(\omega+i\Gamma)^2-M_m^2}\right],
\end{split}
\ee where for brevity we introduced the notation $B\equiv
v_F^2|eB|/c$. The resulting sums are expressed via the digamma
function by means of the formula \be
\sum\limits_{n=0}^\infty\left[\frac{A}{n+a}+\frac{B}{n+b}+\frac{C}{n+c}+\frac{D}{n+d}\right]=
-[A\psi(a)+B\psi(b)+C\psi(c)+D\psi(d)], \ee where for convergence
$A+B+C+D=0$, so that we arrive at the final expressions for the
conductivities (\ref{optical-diagonal}) and (\ref{optical-Hall}). In
the limit of vanishing magnetic field the Hall conductivity becomes
zero, while for the longitudinal conductivity we obtain
\be \label{sigma_ac_B0}
\begin{split}
\sigma_{xx}(\Omega)&=\frac{e^2N_f}{2\pi^2\Omega}{\rm
Re}\int\limits_{-\infty}^\infty
d\omega[n_F(\omega)-n_F(\omega+\Omega)]\left[\left(
\frac{(\omega+i\Gamma)(\omega^\prime
+i\Gamma^\prime)-\Delta^2}{[\omega-\omega^\prime+i(\Gamma-\Gamma^\prime)]
[\omega+\omega^\prime+i(\Gamma+\Gamma^\prime)]}\right.\right.\\
&-\left.\left.\frac{(\omega+i\Gamma)(\omega^\prime-
i\Gamma^\prime)-\Delta^2}{[\omega-\omega^\prime+i(\Gamma+\Gamma^\prime)]
[\omega+\omega^\prime+i(\Gamma-\Gamma^\prime)]}\right)\ln[{\Delta^2-(\omega+i\Gamma)^2}]+
(\omega\leftrightarrow\omega^\prime,
\Gamma\leftrightarrow\Gamma^\prime)\right].
\end{split}
\ee
Calculating the real part we can represent the last expression in
the form
\be
\sigma_{xx}(\Omega)=\frac{e^2N_f}{2\pi^2}\int\limits_{-\infty}^\infty
d\omega\left[\frac{n_F(\omega)-n_F(\omega+\Omega)}{\Omega}\right]{{\cal
A}_L(\omega,\Omega,\Gamma,\Delta)}, \ee
where
\be
\begin{split}
{\cal
A}_L(\omega,\Omega,\Gamma,\Delta)=\frac{1}{D(\omega,\omega^\prime)}&
\left[a\ln\frac{(\Delta^2+\Gamma^{\prime2}
-\omega^{\prime2})^2+4\omega^{\prime2}\Gamma^{\prime2}}{(\Delta^2+\Gamma^2
-\omega^2)^2+4\omega^2\Gamma^2}\right.\\
& \left.+b\arctan\frac{2\omega\Gamma}
{\Delta^2+\Gamma^{2}-\omega^{2}}+c\arctan\frac{2\omega^\prime\Gamma^\prime}
{\Delta^2+\Gamma^{\prime2}-\omega^{\prime2}}\right],
\end{split}
\ee
and \be
\begin{split}
D&(\omega,\omega^\prime)=
[(\Gamma-\Gamma^\prime)^2+(\omega-\omega^\prime)^2][(\Gamma+\Gamma^\prime)^2
+(\omega-\omega^\prime)^2][(\Gamma-\Gamma^\prime)^2+(\omega+\omega^\prime)^2]
[(\Gamma+\Gamma^\prime)^2+(\omega+\omega^\prime)^2],\\
a=&{\Gamma\Gamma^\prime}\left[(\omega^{\prime2}
-\omega^2+\Gamma^{\prime2}-\Gamma^2)[(\omega^{\prime2}+\omega^2+\Gamma^{\prime2}+\Gamma^2)^2
+4(\omega^{\prime2}\omega^2-\Gamma^{\prime2}\Gamma^2)]-8\Delta^2\omega\omega^\prime
(\omega^{\prime2}-\omega^2-\Gamma^{\prime2}+\Gamma^2)\right],\\
b=&2\left[\omega\Gamma^\prime(\omega^{\prime2}
+\omega^2+\Gamma^{\prime2}+\Gamma^2)[(\omega^{\prime2}-\omega^2)^2
+(\Gamma^{\prime2}-\Gamma^2)^2+2(\omega^{\prime2}+\omega^2)
(\Gamma^{\prime2}+\Gamma^2)-8\omega^{\prime2}\Gamma^{2}]\right.\\
& \left.-2\Delta^2\omega^{\prime}
\Gamma^{\prime}[(\omega^{\prime2}-\omega^2)^2
+(\Gamma^{\prime2}-\Gamma^2)^2+2(\omega^{\prime2}+\omega^2)
(\Gamma^{\prime2}+\Gamma^2)-8\omega^{2}\Gamma^2]\right],\\
c=&2\left[\omega^\prime\Gamma(\omega^{\prime2}
+\omega^2+\Gamma^{\prime2}+\Gamma^2)[(\omega^{\prime2}-\omega^2)^2
+(\Gamma^{\prime2}-\Gamma^2)^2+2(\omega^{\prime2}+\omega^2)
(\Gamma^{\prime2}+\Gamma^2)-8\omega^{2}\Gamma^{\prime2}]\right.\\
& \left.-2\Delta^2\omega \Gamma[(\omega^{\prime2}-\omega^2)^2
+(\Gamma^{\prime2}-\Gamma^2)^2+2(\omega^{\prime2}+\omega^2)
(\Gamma^{\prime2}+\Gamma^2)-8\omega^{\prime2}\Gamma^{\prime2}]\right].
\end{split}
\ee
For $\Delta=0$ these expressions are simplified:
\be
\begin{split}
&a={\Gamma\Gamma^\prime}(\omega^{\prime2}-\omega^2+\Gamma^{\prime2}-
\Gamma^2)[(\omega^{\prime2}+\omega^2+\Gamma^{\prime2}+\Gamma^2)^2
+4(\omega^{\prime2}\omega^2-\Gamma^{\prime2}\Gamma^2)],\\
&b=2\left[\omega\Gamma^\prime(\omega^{\prime2}
+\omega^2+\Gamma^{\prime2}+\Gamma^2)[(\omega^{\prime2}-\omega^2)^2
+(\Gamma^{\prime2}-\Gamma^2)^2+2(\omega^{\prime2}+\omega^2)
(\Gamma^{\prime2}+\Gamma^2)-8\omega^{\prime2}\Gamma^{2}]\right],\\
&c=2\left[\omega^\prime\Gamma(\omega^{\prime2}
+\omega^2+\Gamma^{\prime2}+\Gamma^2)[(\omega^{\prime2}-\omega^2)^2
+(\Gamma^{\prime2}-\Gamma^2)^2+2(\omega^{\prime2}+\omega^2)
(\Gamma^{\prime2}+\Gamma^2)-8\omega^{2}\Gamma^{\prime2}]\right].
\end{split}
\ee
The behavior of $\sigma_{xx}(\Omega)$ in this limit was studied in
Ref.~\cite{Ando2002JPSJ} for graphene and in Ref.~\cite{Kim2004PRB}
for a $d$-wave superconductor. For $B=0$ the latter is rather
similar to graphene at $\mu=0$.

\section{Derivation of the Hall conductivity
in the clean limit from E\lowercase{q.}~(\ref{sigma_xy_final}) }
\label{sec:B}

Eq.~(\ref{hall_cond}) can be obtained directly from
Eq.~(\ref{sigma_xy_final}) . Indeed taking the limit $\Gamma\to0$,
we find from Eq.~(\ref{AH}) that
\be \mathcal{A}_{H}(\omega, B, \Gamma, \Delta) = \frac{1}{2
\pi^2}\left[\arctan\frac{2\omega\Gamma}
{\Delta^2+\Gamma^2-\omega^2}-2{\rm
Im}\ln\Gamma\left(\frac{\Delta^2+\Gamma^2
-\omega^2-2i\omega\Gamma}{B}\right)\right],\quad \Gamma\to0. \ee
Recall that here $B$ is a short-hand notation for $v_F^2|eB|/c$. We
begin with the expression
\begin{equation}
\begin{split}
&{\rm
Im}\ln\Gamma\left(\frac{\Delta^2+\Gamma^2-\omega^2-2i\omega\Gamma}{2B}\right)
={\rm sgn}\,\omega{\rm
Im}\ln\Gamma\left(\frac{\Delta^2+\Gamma^2-\omega^2
-2i|\omega|\Gamma}{2B}\right)\\
&={\rm sgn}\,\omega{\rm Im}\ln\Gamma\left(\frac{\Delta^2
+\Gamma^2-\omega^2-2i|\omega|\Gamma}{2B}\right)\left[\theta(\Delta^2+\Gamma^2
-\omega^2)+\theta(\omega^2-\Delta^2-\Gamma^2)\right].
\end{split}
\end{equation}
Now we use the relationship
\be \Gamma(z)\Gamma(-z)=\frac{\pi}{z\sin(-\pi z)} \ee
to rewrite the last expression in the form
\begin{equation}
\begin{split}
&{\rm
Im}\ln\Gamma\left(\frac{\Delta^2+\Gamma^2-\omega^2-2i\omega\Gamma}{2B}\right)
=-{\rm sgn}\,\omega\left\{{\rm
Im}\ln\Gamma\left(\frac{|\omega^2-\Delta^2
-\Gamma^2|+2i|\omega|\Gamma}{2B}\right)\right.\\
&\left.+\theta(\omega^2-\Delta^2-\Gamma^2)
\left[\pi-\arctan\frac{2|\omega|\Gamma}{|\omega^2-\Delta^2-\Gamma^2|}
+{\rm
Im}\ln\sin\left(\pi\frac{\omega^2-\Delta^2-\Gamma^2+2i|\omega|
\Gamma}{2B}\right)\right]\right\}.
\end{split}
\end{equation}
Hence \begin{equation}
\begin{split}
&\mathcal{A}_H(\omega, B, \Gamma, \Delta) = \frac{{\rm sgn}
\,\omega}{2 \pi^2} \left\{
\arctan\frac{2|\omega|\Gamma}{|\omega^2-\Delta^2-\Gamma^2|}+2{\rm
Im}\ln\Gamma
\left(\frac{|\omega^2-\Delta^2-\Gamma^2|+2i|\omega|\Gamma}{2B}\right)\right.
\\&\left.-\theta(\omega^2-\Delta^2-\Gamma^2)\left[\pi-2{\rm
Im}\ln\sin\left(\pi \frac{\omega^2-\Delta^2-\Gamma^2+2i|\omega|
\Gamma}{2B}\right)\right]\right\}.
\end{split}
\end{equation}
Now using the formula
\be {\rm
Im}\ln\sin(a+ib)=\frac{\pi}{2}-a-\sum\limits_{k=1}^\infty\frac{1}{k}\sin(2ka)
e^{-2kb},\quad b>0, \ee
we arrive at the following representation
\be
\begin{split} & \mathcal{A}_H(\omega, B, \Gamma, \Delta) = \frac{{\rm
sgn}\,\omega}{2 \pi^2}\left\{
\arctan\frac{2|\omega|\Gamma}{|\omega^2-\Delta^2-\Gamma^2|}+2{\rm
Im}\ln\Gamma
\left(\frac{|\omega^2-\Delta^2-\Gamma^2|+2i|\omega|\Gamma}{2B}\right)\right.\\
&\left.
-\pi\theta(\omega^2-\Delta^2-\Gamma^2)\left[\frac{\omega^2-\Delta^2-\Gamma^2}{B}
+2\sum\limits_{k=1}^\infty\frac{1}{\pi k}\sin\left(2\pi
k\frac{\omega^2-\Delta^2
-\Gamma^2}{2B}\right)e^{-\frac{2k|\omega|\Gamma}{B}}\right]\right\}.
\end{split}
\ee
Finally by means of the identity (4.6) of
Ref.~\cite{Sharapov2004PRB}, we obtain in the limit $\Gamma\to0$
that
\begin{equation}
\begin{split} &\mathcal{A}_H(\omega, B, \Gamma,
\Delta) =-\frac{1}{2\pi}{\rm sgn}\,\omega\theta(\omega^2-\Delta^2)
\left[\frac{\omega^2-\Delta^2}{B}
+2\sum\limits_{k=1}^\infty\frac{1}{\pi k}\sin\left(2\pi
k\frac{\omega^2-\Delta^2}
{2B}\right)\right]\\
&=-\frac{1}{2\pi}{\rm
sgn}\,\omega\left[\theta(\omega^2-\Delta^2)+2\sum\limits_{n=1}^\infty\theta(\omega^2
-\Delta^2-2Bn)\right].
\end{split}
\end{equation}
Inserting the last expression in Eq.~(\ref{sigma_xy_final}) and
integrating over $\omega$ we finally arrive at
Eq.~({\ref{hall_cond}}).

\section{The equation for chemical potential} \label{sec:D}

To derive the relationship  for the carrier imbalance $\rho$
($\rho\equiv n_e - n_h$, where $n_e$ and $n_h$ are the densities of
``electrons'' and ``holes'', respectively) and the chemical
potential $\mu$, we begin with the well-known expression
\begin{equation}
\label{GF2rho} \rho =  \mbox{tr} [\gamma^0 \tilde{S}(\tau,
\mathbf{0})], \quad \tau \to 0,
\end{equation}
where $\tilde{S}(\tau,\mathbf{r})$ is the translation invariant
part of the Green's function (\ref{Green}).  Taking its Fourier
transform and using the spectral representation
(\ref{spectr_repr}), we arrive at
\begin{equation}
\label{rho-exp} \rho=
T\sum\limits_{n=-\infty}^\infty\int\frac{d^2k}{(2\pi)^2}\int_{-\infty}^\infty
d\omega\frac{{\rm tr}[\gamma^0A(\omega,{\bf k})]}
{i\omega_n+\mu-\omega}.
\end{equation}
After evaluating the sum over Matsubara frequencies we obtain
\begin{equation}
\rho=-\frac{1}{2}\int\frac{d^2k}{(2\pi)^2}\int_{-\infty}^\infty
d\omega\tanh\frac{\omega-\mu}{2T} {\rm tr}[\gamma^0A(\omega,{\bf
k})].
\end{equation}
Now taking into account that ${\rm tr}[\gamma^0A(\omega,{\bf k})]$
is an even function of $\omega$,  we may write
\begin{equation}
\begin{split} \rho=&\frac{1}{2}\int\frac{d^2k}{(2\pi)^2}\int_{0}^\infty
d\omega\left[\tanh\frac{\omega+\mu}{2T}-\tanh\frac{\omega-\mu}{2T}\right]
{\rm tr}[\gamma^0A(\omega,{\bf k})]\\
=&\int\frac{d^2k}{(2\pi)^2}\int_{0}^\infty
d\omega\left[n_F(\omega-\mu)-n_F(\omega+\mu)\right] {\rm
tr}[\gamma^0A(\omega,{\bf k})].
\end{split}
\end{equation}
Substituting  the spectral function (\ref{spectral-function}) in the
first line of the last equation and integrating over momenta, we
arrive at
\begin{equation}
\label{n-Gamma}
\rho=\frac{N_f|eB|}{4\pi^2c}\sum\limits_{n=0}^\infty\alpha_n\int_{-\infty}^\infty
d\omega\tanh\frac{\omega+\mu}{2T}\left(\frac{\Gamma}{(\omega-M_n)^2+\Gamma^2}
+(\omega\to-\omega)\right),\quad \alpha_0=1,\quad \alpha_n=2,\,
n\ge1.
\end{equation}
The ratio $\alpha_n/\alpha_0 =2$ for $n\ge 1$ is related to the
above-mentioned smaller degeneracy of the $n=0$ Landau level. It
is easy to see that Eq.~(\ref{n-Gamma}) in the limit $\Gamma=0$
reduces to
\begin{equation}
\label{mu-n:eq} \rho=\frac{N_f|eB|}{4\pi c}\left[\tanh
\frac{\mu+\Delta}{2T}+\tanh\frac{\mu-\Delta}{2T}+2\sum\limits_{n=1}^\infty
\left(\tanh\frac{\mu+M_n}{2T}+\tanh\frac{\mu-M_n}{2T}\right)\right],
\end{equation}
while in the limit $B=\Gamma=T=\Delta=0$ it reduces to
Eq.~(\ref{number}) (see, for example, Eq.~(77) in
Ref.~\cite{Gorbar2002PRB}) giving the relationship between the
chemical potential and the free carrier imbalance.

Comparing Eq.~(\ref{mu-n:eq}) with Eq.~(\ref{hall_cond}) we
finally obtain that in the ideally clean system (see also
Ref.~\cite{Gorbar2002PRB})
\begin{equation}
\label{hall-clean} \sigma_{xy}=-\frac{ec\rho}{B}.
\end{equation}
The last expression seems to be paradoxical at first glance, because
it corresponds to the classical expression (\ref{classics}) far
beyond the validity of the classical limit (see
Sec.~\ref{sec:Drude-Zener}). Nevertheless this result is absolutely
correct and it shows the consistency of our calculation. As
explained in Ref.~\cite{Girvin-lectures}, Eq.~(\ref{hall-clean}) is
expected for an ideal conductor. This similarity between
Eq.~(\ref{mu-n:eq})  and Eq.~(\ref{hall_cond}) was exploited in
Ref.~\cite{Schakel1991PRD}, where instead of calculating the
electrical conductivity $\sigma_{xy}$, the density (\ref{mu-n:eq})
was obtained. However, to consider the IQHE one must take into
account the presence of impurities
\cite{Hajdu.book,Girvin-lectures,Prange-book}. It is believed that
they lead to the localization of most of the bulk states, except in
a region around the center of the Landau band, and act as a
reservoir which almost fixes the chemical potential
\cite{Hajdu.book,Prange-book}. It turns out that in this more
physical situation it still makes sense to rely on
Eq.~(\ref{hall_cond}) for $\sigma_{xy}$, while Eq.~(\ref{mu-n:eq})
cannot be used in the IQHE regime. One particular model of the
equation for $\mu$ that includes the reservoir could be a
generalization of the one discussed in Ref.~\cite{Grigoriev2001JETP}
for nonrelativistic quasiparticles.

The main implication of this model is that the density of the
delocalized carriers in the Hall bar may oscillate as the field $B$
varies. It seems the oscillations of this kind were indeed observed
in the IQHE system \cite{Raymond1999ApplSurf}. However, there is no
consensus on a microscopic picture of the localization in the
quantum Hall effect (see e.g.
Refs.~\cite{Huckestein1995RMP,Ilani2004Nature,Steele2005PRL}) even
in 2D electron gas. On the other hand, the localization of the Dirac
quasiparticles appears to be quite different from the localization
in 2D electron gas with the parabolic dispersion \cite{Geim-new}.
This indicates that further studies of the localization and
oscillations of the density of the delocalized carriers in graphene
may be very useful.

\section{Solution of the Dirac equation in the symmetric gauge} \label{sec:E}

The Dirac equation in the problem of a relativistic fermion in a
constant magnetic field $B$ takes the following form in $2+1$
dimensions:
\be \label{eq:fre}\left[i\tilde{\gamma}^0\hbar\partial_t + i v_F
\tilde{\gamma}^1 \left(\hbar \partial_x + i
\frac{e}{c}A_x^{\mathrm{ext}}\right) + i v_F \tilde{\gamma}^2
\left(\hbar
\partial_y + i \frac{e}{c}A_y^{\mathrm{ext}}\right) - \Delta \right]\psi(t,{\mathbf
r})=0, \ee
where the vector potential $\mathbf{A}^{ext}=(-By/2, Bx/2)$, so
that the magnetic field $\mathbf{B}= \nabla \times
\mathbf{A}^{\mathrm{ext}}$ is directed along the positive z axis.
In $2+1$ dimensions, there are two inequivalent representations of
the Dirac algebra (see e.g.,Ref.~\cite{Jackiw1981PRD}):
\be \label{eq:pauli} \tilde{\gamma}^0=\sigma_3,
\tilde{\gamma}^1=i\sigma_1, \tilde{\gamma}^2=i\sigma_2 \ee and \be
\label{eq:pauli1} \tilde{\gamma}^0=-\sigma_3,
\tilde{\gamma}^1=-i\sigma_1, \tilde{\gamma}^2=-i\sigma_2. \ee
which correspond to right- and left-handed coordinate systems. Here
$\sigma_i$ are Pauli matrices. The representation of gamma matrices
$(\sigma_3, i\sigma_2,-i\sigma_1)$ used to write the Lagrangian
(\ref{Lagrangian}) is related to the representation $(\sigma_3,
i\sigma_1,i\sigma_2)$ by means of the unitary transformation
$U=(\hat{I}+i\sigma_3)/\sqrt{2}$. Although the final results of the
calculations do not depend on either the representation of
$\gamma$-matrices or the gauge, the intermediate expressions depend
on this choice.

Since the representation (\ref{eq:pauli}), (\ref{eq:pauli1}) is
more commonly used in the literature, in this Appendix we solve
the Dirac equation (\ref{eq:fre}) and obtain the operator
(\ref{pseudospin}) using this representation.

Let us begin by considering the representation (\ref{eq:pauli}). The
energy spectrum in the problem (\ref{eq:fre}) depends on the sign of
$eB$; let us for definiteness assume that $eB>0$. Then, the energy
spectrum takes the form (to be concrete, we assume also that
$\Delta\geq 0$):
\be \label{eq:5}
\begin{split}
E_0 &= -M_0=-\Delta, \\
E_n &= \pm M_n=\pm\sqrt{\Delta^2+2n{|eB|\hbar v_F^2}/{c}},\quad
n=1,2,\dots
\end{split}
\ee
(the Landau levels). The general solution is
\be \label{eq:sol} \psi(x)=\sum_{n,m}\left[ a_{nm}u_{nm}(x) +
b^+_{nm} v_{nm}(x)\right], \ee
where
\be \label{solution_positive}
\begin{split}
& u_{nm}=\frac{1}{l\sqrt{2\pi}}\exp (-iM_nt)\frac{1}{\sqrt{2M_n}}
\left(\begin{array}{cc}\sqrt{M_n+\Delta}\,
J^{n-m}_{m-1}(\xi)e^{i(m-1)\theta} \\
\sqrt{M_n-\Delta}\,J^{n-m}_{m}(\xi)e^{im\theta}
\end{array}\right),\quad n\ge1,\quad m\leq n,\\
& v_{0m}=\frac{1}{l\sqrt{2\pi}}\exp (iM_0t)\left(\begin{array}{cc}0
 \\
J^{-m}_{m}(\xi)e^{im\theta}   \end{array}\right),\quad n=0,\quad
m\leq
0, \\
& v_{nm}=\frac{1}{l\sqrt{2\pi}}\exp (iM_nt)\frac{1}{\sqrt{2M_n}}
\left(\begin{array}{cc}-\sqrt{M_n-\Delta}\,
J^{n-m}_{m-1}(\xi)e^{i(m-1)\theta} \\
\sqrt{M_n+\Delta}\,J^{n-m}_{m}(\xi)e^{im\theta}
\end{array}\right),\quad n\ge 1,\quad m\leq n.
\end{split}
\ee
Here the functions \cite{Melrose}
\be
J^n_\nu(\xi)=\left[\frac{n!}{(n+\nu)!}\right]^{1/2}e^{-\xi/2}\xi^{\nu/2}L_n^\nu(\xi),
\qquad J_m^{-m}(\xi)=\frac{(-1)^m}{\sqrt{(-m)!}}e^{-\xi/2}\xi^{-
m/2} \quad (m\leq 0),\ee
where $L_n^m(\xi)$ are Laguerre polynomials ($L_{n}^m(\xi)\equiv0$
for $n\leq -1$), $l\equiv(\hbar c/|eB|)^{1/2}$ is the magnetic
length, $\xi={\mathbf r}^2/2l^2$, the quantum number $m\leq n$
reflects the degeneracy of each level in the angular momentum. The
spinors $u_{nm}$ and $v_{nm}$ are normalized as follows:
\be \int\,d^2x\,u_{n^\prime m^\prime}^\dagger(x) u_{nm}(x)=
\int\,d^2x\,v_{n^\prime m^\prime}^\dagger(x)
v_{nm}(x)=\delta_{n^\prime,n}\delta_{m^\prime,m}. \ee
Thus the lowest Landau level with $n=0$ is special: while at
$n\geq1$, there are solutions corresponding to both fermion
$(E_n=M_n)$ and antifermion $(E_n=-M_n)$ states, the solution with
$n=0$ describes only antifermion states.

If we used the representation (\ref{eq:pauli1}) for Dirac's
matrices, the general solution would be given by Eq.~(\ref{eq:sol})
with $u_{nm}(x)$, $v_{nm}(x)$ being substituted by
$v_{nm}(-x),u_{nm}(-x)$:
\be \label{eq:9}
\psi(x)=\sum\limits_{n,m}\left[c_{nm}v_{nm}(-x)+d_{nm}^\dagger
u_{nm}(-x)\right]. \ee
The solution $v_{0m}$ corresponds to the Landau level $n=0$ with
positive energy $E_0=\Delta$, while the solutions $v_{nm},u_{nm}$
with $n\ge1$ correspond to the energy eigenvalues $E_n=\pm M_n$
(compare with Eq.~(D4)). Accordingly, when the spectra for two
inequivalent representations are united together the degeneracy of
the $n=0$ level turns out to be a half of the degeneracy of the
levels with $n\geq1$. The four-component spinor $\Psi$ is composed
from two two-component spinors Eqs.~(\ref{eq:sol}) and (\ref{eq:9})
and using it one can obtain the expression (\ref{pseudospin}) for
the pseudospin rotation generator.

\newpage

\begin{thebibliography}{99}

\bibitem{Novoselov2004Science} K.S.~Novoselov, A.K.~Geim, S.V.~Morozov,
D.~Jiang, Y.~Zhang, S.V.~Dubonos, I.V.~Grigorieva, and A.A.~Firsov,
Science {\bf 306}, 666 (2004); K.S.~Novoselov, A.K.~Geim,
S.V.~Morozov, S.V.~Dubonos, Y.~Zhang, and D.~Jiang,
cond-mat/0410631.

\bibitem{Novoselov2005PNAS} K.S.~Novoselov, D.~Jiang, T.~Booth, V.V.~Khotkevich,
S.M.~Morozov, A.K.~Geim, Proc.~Nat.~Acad.~Sc. {\bf 102}, 10451
(2005); see also the latest work on bilayer graphene,
K.S.~Novoselov, E.~McCann, S.V.~Morozov, V.I.~Fal'ko,
M.I.~Katsnelson, U.~Zeitler, D.~Jiang, F.~Schedin and A.~K. Geim,
Nature Physics {\bf 22}, 177 (2006).

\bibitem{Berger2004JPCB} C.~Berger, Z.~Song, T.~Li, X.~Li, A.Y.~Ogbazghi,
R.~Feng, Z.~Dai, A.N.~Marchenkov, E.H.~Conrad, P.N.~First, and
W.A.~de Heer, J. Phys. Chem. B {\bf 108}, 19912 (2004).

\bibitem{Zhang2005PRL} Y.~Zhang, J.P.~Small, M.E.S.~Amori, and
P.~Kim,  \prl {\bf 94}, 176803 (2005).

\bibitem{Bunch2005Nano} J.S.~Bunch, Y.~Yaish, M.~Brink, K.~Bolotin, and
P.L.~McEuen, Nano Letters {\bf 5}, 287 (2005).

\bibitem{Morozov2005}
S.V.~Morozov, K.S.~Novoselov, F.~Schedin, D.~Jiang, A.A.~Firsov, and
A.K.~Geim, \prb {\bf 72}, 201401(R) (2005).

\bibitem{Geim-new}
K.S.~Novoselov, A.K.~Geim, S.V.~Morozov, D.~Jiang, M.I.~Katsnelson,
I.V.~Grigorieva, S.V.~Dubonos, A.A.~Firsov, Nature {\bf 438}, 197
(2005).

\bibitem{Kim-new} Y.~Zhang, Y.-W.~Tan, H.L.~Stormer, P.~Kim, Nature {\bf 438}, 201
(2005).

\bibitem{Semenoff1984PRL} G.W.~Semenoff,
\prl {\bf 53}, 2449 (1984).

\bibitem{Gonzales1993NP} J.~Gonz{\`a}lez, F.~Guinea, and M.A.H.~Vozmediano,
Nucl.~Phys. B {\bf 406}, 771 (1993).

\bibitem{Gusynin2005PRL} V.P.~Gusynin and S.G.~Sharapov,
\prl {\bf 95}, 146801 (2005).

\bibitem{Peres2005} N.M.R.~Peres, F.~Guinea, A.H.~Castro~Neto,
preprint cond-mat/0506709.

\bibitem{Dresselhaus1974PRB}  G.~Dresselhaus,
Phys. Rev. B {\bf 10}, 3602 (1974).

\bibitem{Uji1998PB} S.~Uji, J.S.~Brooks, and Y.~Iye,
Physica B {\bf 246-247}, 299 (1998).

\bibitem{Kopelevich2003PRL} Y.~Kopelevich, J.H.S.~Torres, R.R.~da Silva, F.~Mrowka, H.~Kempa, and
P.~Esquinazi, Phys.~Rev.~Lett. {\bf 90}, 156402 (2003).

\bibitem{Kempa2006SSC} H.~Kempa, P.~Esquinazi, Y.~Kopelevich,
Solid State Commun. {\bf 138}, 118 (2006).

\bibitem{Ocana2003PRB}
R.~Oca{\~n}a, P.~Esquinazi, H.~Kempa, J.H.S.~Torres, and
Y.~Kopelevich, \prb {\bf 68}, 165408 (2003).

\bibitem{Sharapov2004PRB} S.G.~Sharapov, V.P.~Gusynin, and H.~Beck,
Phys. Rev. B {\bf 69}, 075104 (2004).

\bibitem{Gusynin2005PRB} V.P.~Gusynin and S.G.~Sharapov,
Phys. Rev. B {\bf 71}, 125124 (2005).

\bibitem{Luk'yanchuk2004PRL} I.A.~Luk'yanchuk and Y.~Kopelevich,
\prl {\bf 93}, 166402 (2004).

\bibitem{Matsui2005PRL} T.~Matsui, H.~Kambara, Y.~Niimi, K.~Tagami, M.~Tsukada, and H.~Fukuyama
\prl {\bf 94}, 226403 (2005).

\bibitem{Li2005}
Z.~Li, W.~Padilla, S.~Dordevic, P.~Esquinazi, C.C.~Homes, D.~Basov,
Abstracts of March 2005 APS Meeting.

\bibitem{Nersesyan1989JLTP} A.A.~Nersesyan and G.E.~Vachanadze, J.~Low~Temp.~Phys. {\bf 77}, 293
(1989); A.A.~Nersesyan, G.I.~Japaridze, and G.E. Vachanadze,
J.~Phys.~Cond.~Matt. {\bf 3}, 3353 (1991).

\bibitem{Beaugnon1991Nature} E.~Beaugnon and R. Tournier, Nature {\bf 349},
470 (1991).

\bibitem{Gorbar2002PRB} E.V.~Gorbar, V.P.~Gusynin, V.A.~Miransky, and I.A.~Shovkovy,
Phys. Rev. B {\bf 66}, 045108 (2002).

\bibitem{Sharapov2003PRB}
S.G.~Sharapov, V.P.~Gusynin, and H.~Beck, Phys. Rev. B {\bf 67},
144509 (2003).

\bibitem{Zhang2006} Y.~Zhang, Z.~Jiang, J.P.~Small, M.S.~Purewal,
Y.-W.~Tan, M.~Fazlollahi, J.D.~Chudow, J.A.~Jaszczak, H.L.~Stormer,
P. Kim, Phys. Rev. Lett. {\bf 96}, 136806 (2006).

\bibitem{Saito.book} R.~Saito, G.~Dresselhaus and M.S.~Dresselhaus,
{\em Physical Properties of Carbon Nanotubes} (Imperial College
Press, London, 1998).

\bibitem{Khveshchenko2001PRL}  D.V.~Khveshchenko,
Phys.~Rev.~Lett. {\bf 87}, 246802 (2001); D.V.~Khveshchenko and
H.~Leal, Nucl.~Phys.~B {\bf 687}, 323 (2004).

\bibitem{Gusynin1995PRD} V.P.~Gusynin, V.A.~Miransky, and I.A.~Shovkovy, Phys.~Rev.~Lett.
{\bf 73}, 3499 (1994);
Phys.~Rev.~D {\bf 52}, 4718 (1995).

\bibitem{Chodos1990PRD}
A.~Chodos, K.Everding, and D.A.~Owen, Phys.~Rev.~D {\bf 42}, 2881
(1990).

\bibitem{Zheng2002PRB} Y.~Zheng and T.~Ando, Phys. Rev. B {\bf 65}, 245420 (2002).

\bibitem{Champel2002PRB} T.~Champel and V.P.~Mineev, Phys.~Rev.~B
{\bf 66}, 195111 (2002);  {\bf 67}, 089901(E) (2003).

\bibitem{Ferrer2003EPJB} E.J.~Ferrer, V.P.~Gusynin, and V.~de~la~Incera,
Eur.~Phys.~J. B {\bf 33}, 397 (2003).

\bibitem{Shon1998JPSJ} N.H.~Shon and T.~Ando, J.Phys.~Soc.~Jpn.
{\bf 67}, 2421 (1998).

\bibitem{foot3}
This universality has the same origin as the universality considered
in E.~Fradkin, Phys.~Rev.~B {\bf 33}, 3263 (1986) for degenerate
semiconductors and in P.A.~Lee, Phys. Rev. Lett. {\bf 71}, 1887
(1993) for $d$-wave superconductors in the absence of magnetic
field.

\bibitem{Suzuura2002} H.~Suzuura and T.~Ando, {\em Physics of Semiconductors 2002}, edited by A.
R.~Long and J.H.~Davies (Institute of Physics
    Publishing, Bristol, 2003).

\bibitem{Jonson1984PRB} M.~Jonson and S.~M.~Girvin, Phys. Rev. B {\bf 29}, 1939 (1984).

\bibitem{foot1} Since the Zeeman term is considered to be small,
our definition of the filling factor $\nu_B = 2\pi \hbar
c|\rho|/(N_f |eB|)$ also counts the spin degeneracy, so that for
$N_f=2$ it is half of the normally used \cite{Hajdu.book} filling
factor, $\nu= 2 \pi \hbar c|\rho|/|eB|$.

\bibitem{Hajdu.book}  M.~Jan$\beta$en, O.~Veihweger, U.~Fastenrath, and J.~Hajdu,
{\em Introduction to the Theory of the Integer Quantum Hall Effect},
edited by J.~Hajdu (VCH, Weinheim, 1994).

\bibitem{MacDonald1983PRB} A.H.~MacDonald, Phys. Rev. B {\bf 28}, 2235 (1983).

\bibitem{Nieto1985AJP} M.M.~Nieto and P.L.~Taylor,
Am.~J.~Phys. {\bf 53}, 234 (1985).

\bibitem{Haldane1988PRL} F.D.M.~Haldane, \prl {\bf 61}, 2015 (1988).

\bibitem{Schakel1991PRD} A.M.J.~Schakel, Phys.~Rev.~D {\bf 43}, 1428
(1991).

\bibitem{Abouelsaood1985PRL} A.~Abouelsaood, \prl {\bf 54}, 1973
(1985).

\bibitem{Smrcka1997JPC} L.~Smr{\v c}ka and P.~St{\v r}eda,
J.~Phys.~C {\bf 10}, 2153 (1977).

\bibitem{Oji1985PRB} H.~Oji and P.~Streda, Phys.~Rev. B {\bf 31}, 7291 (1985).

\bibitem{Cooper1997PRB} N.R. Cooper, B. I. Halperin, and I. M. Ruzin, Phys. Rev. B 55,
2344 (1997).

\bibitem{Oganesyan2004PRB} V.~Oganesyan and I.~Ussishkin, Phys. Rev. B {\bf 70}, 054503 (2004).

\bibitem{Dora2003PRB}
B.~D\'ora, K.~Maki, A.V\'anyolos, and A.~Virosztek, Phys.~Rev.~B
{\bf 68}, 241102(R) (2003).

\bibitem{Prange-book} {\em The Quantum Hall Effect}, edited by R.E.~Prange and
S.M.~Girvin (Springer-Verlag, New York, 1987).

\bibitem{Ando2002JPSJ} T.~Ando, Y.~Zheng, and H.Suzura, J.Phys.~Soc.~Jpn.
{\bf 71}, 1318 (2002).

\bibitem{Kim2004PRB}
W.~Kim, F.~Marsiglio, and J.P.~Carbotte, Phys.~Rev.~B {\bf 70},
060505(R) (2004).

\bibitem{Girvin-lectures} S.M.~Girvin, {\em The Quantum Hall Effect: Novel Excitations and Broken
Symmetries}, in Topological Aspects of Low Dimensional Systems, ed.
A.~Comtet, T.~Jolicoeur, S.~Ouvry, F.~David (Springer-Verlag, Berlin
and Les Editions de Physique, Les Ulis, 2000).


\bibitem{Grigoriev2001JETP} P.~Grigoriev, JETP {\bf 92}, 1090 (2001)
[ZhETF {\bf 119}, 1257 (2001)].

\bibitem{Raymond1999ApplSurf} A.~Raymond, S.~Juillaguet, I.~Elmezouar, W.~Zawadzki,
M.L.~Sadowski, M.~Kamal-Saadik, and B. Etienne, Semicond. Sci.
Technol. {\bf 14}, 915 (1999).

\bibitem{Huckestein1995RMP} B.~Huckestein, Rev.~Mod.~Phys. {\bf
67}, 357 (1995).

\bibitem{Ilani2004Nature} S.~Ilani, J.~Martin, E.~Teitelbaum,
J.H.~Smet, D.~Mahalu, V.~Umansky, and A.~Yacoby, Nature {\bf 427},
328 (2004).

\bibitem{Steele2005PRL} G.A.~Steele, R.C.~Ashoori, L.N.~Pfeiffer,
and K.W.~West, \prl {\bf 95}, 136804 (2005).

\bibitem{Jackiw1981PRD}
R.~Jackiw and S.~Templeton, Phys. Rev. D {\bf 23}, 2291 (1981).

\bibitem{Melrose} D.B.~Melrose and A.J.~Parle, Aust.~J.~Phys. {\bf 36}, 755 (1983).

\end{thebibliography}
\end{document}